\newcommand{\ApJ}[3]{ #1, {ApJ, \/} {#2}, #3}

\newcommand{\ApJS}[3]{ #1, {ApJS, \/} {#2}, #3}
\newcommand{\AJ}[3]{ #1, {AJ, \/} {#2}, #3}
\newcommand{\AsAp}[3]{ #1, {A\& A, \/} {#2}, #3}

\newcommand{\ARAA}[3]{ #1, {ARA\& A, \/} {#2}, #3}

\newcommand{\PASP}[3]{ #1, {PASP, \/} {#2}, #3}

\def\etal{{\it et al.\/}}
\def\ph{\phantom}
\def\cf{{\it cf.\/}}
\documentstyle[12pt,aaspp4]{article}
\lefthead{Musella, Piotto and Capaccioli}
\righthead{Cepheids in NGC 3109}

\begin{document}

\received{}
\accepted{}
\journalid{}{}
\articleid{}{}
 
\title{On the Cepheid variables of nearby galaxies III. NGC 3109}

\author{Ilaria Musella} 
\affil{Osservatorio Astronomico di Capodimonte,  via Moiariello 16,\\
              I--80131 Napoli,  Italy}
\authoraddr{ilaria@astrna.na.astro.it}
\author{Giampaolo Piotto }
\affil{Dipartimento di Astronomia, Universit\`a di Padova,  
              I--35122 Padova,  Italy}
\authoraddr{piotto@astrpd.pd.astro.it}

\author{Massimo Capaccioli}
\affil{Osservatorio Astronomico di Capodimonte,  salita Moiariello 16,\\
              I--80131 Napoli,  Italy \\ and \\Dipartimento di Scienze 
       Fisiche, Universit\`a di Napoli\\
       Mostra d'Oltremare, Padiglione 19, I--80125--Italy}
\authoraddr{capaccioli@astrna.na.astro.it}

\begin{abstract}

We extended to the R and I bands the light curve coverage for 8 Cepheids
already studied in B and V by Capaccioli \etal\ [AJ, 103, 1151 (1992)].
Sixteen additional Cepheid candidates have been identified and
preliminary periods are proposed. The new Cepheids allow the
period-luminosity relation to be extended one magnitude fainter.
Apparent B, V, R, and I distance moduli have been calculated. Combining
the data at different wavelengths, and assuming a true distance modulus
of 18.50 mag for the LMC, we obtain for NGC~3109 a true distance modulus
$(m-M)_0=25.67\pm0.16$, corresponding to $1.36\pm0.10$ Mpc. Adopting
$E(B-V)=0.08$ for the LMC, the interstellar reddening for the Cepheids
in NGC 3109 is consistent with 0. A discussion on the possible
implications of this result is presented. A comparison of the
period-color, period-amplitude, and period-luminosity relations suggests
similar properties for the Cepheids in the LMC, NGC 3109, Sextans A,
Sextans B, and IC 1613, though the uncertainties in the main parameter
determination are still unsatisfactorily high for a firm conclusion on
the universality of the period-luminosity relation.

\end{abstract}

   \keywords{Cepheids --- galaxies: distances and redshifts --- galaxies:
individual (NGC~3109)}

\section{Introduction} \label{intro} This paper is part of a program
aimed at increasing the number of galaxies with accurate distance
determinations via ground-based multicolor CCD photometry of Cepheid
variables.

In the first paper of the series, Capaccioli, Piotto and Bresolin (1991,
CPB) presented new $BV$ photometry of the Cepheids belonging to the
sample of variables already identified by Sandage and Carlson (1988, SC)
in NGC 3109. In that work, CPB derived a new zero point for the
photometric scale. The new data gave a distance modulus
$\mu_0=25.5$~mag, $\sim25$\% shorter than previously measured by SC. The
availability of only two photometric bands ($B$ and $V$) did not allow
CPB to apply the multicolor method discussed by Freedman (1985, F85) to
directly estimate the internal absorption of the Cepheids in NGC~3109.
Actually, Bresolin, Capaccioli, and Piotto (1993, BCP) pointed out that
the internal absorption in some of the fields studied by CPB could be
greater than the adopted average value E(B-V)=0.04. Besides improving
the photometry of the Cepheids previously discovered on photographic
plates, CCD data allow us also to obtain more accurate and deeper
photometry in crowded fields, enabling the discovery of new fainter
Cepheids. The consequent extension of the period--luminosity relation
($PL$) to shorter periods permits a more accurate determination of its
zero point. For instance, in the second paper of this series, Piotto,
Capaccioli and Pellegrini (1994, PCP) found five new Cepheids in Sextans
A and four in Sextans B, in addition to the five variables in Sextans A
and the three in Sextans B already discovered by Sandage and Carlson
(1982, 1985). The new variables extended the faint end of the $PL$
relation of these two galaxies by one magnitude. PCP observed these two
galaxies in three passbands: $B$, $V$ and $I$, again obtaining a new
zero point for the photometric calibration. Using the multiwavelength
photometry method (F85) they derived a true distance modulus, corrected
for interstellar extinction, of $\mu_0=25.71$ for Sextans A and of
$\mu_0=25.63$ for Sextans~B, assuming a distance modulus $\mu_0=18.50$
for the Large Magellanic Cloud (LMC).

In the framework of this program, here we present a new determination of
the distance to the galaxy NGC~3109 based on a new $BVRI$ photometry of
the Cepheids located in the field F1 of BCP. The data set and the
reduction procedures are illustrated in the Section 2. In Section 3 we
present the already known variables and our 16 new Cepheid candidates on
which the present distance determination of NGC 3109 is based. The
determination of the Cepheid parameters is discussed in Section 4, while
Section 5 is devoted to the determination of the distance to NGC 3109. A
comparison of the PL and period-color relations for the Cepheids in a
few nearby galaxies is presented in Section 6. A brief discussion and a
comparison with the previous determinations of the distance to NGC 3109
is in Section 7.

\section{Observations and Reductions}

This study is a follow up of the work by CPB. In that paper it was
impossible to identify new Cepheids due to the limited time coverage and
small number of data points (six at most). We then collected, in December
1991, March 1992 and February 1993, a series of $B$, $V$, $R$, and
$I$--band images of a $1'.9\times3'.0$ field (Fig. 1) of NGC~3109 coded
as F1 (\cf\ Table~\ref{table1} for the log of the observations).

\begin{figure} \figurenum{1} \epsscale{0.60} \plotone{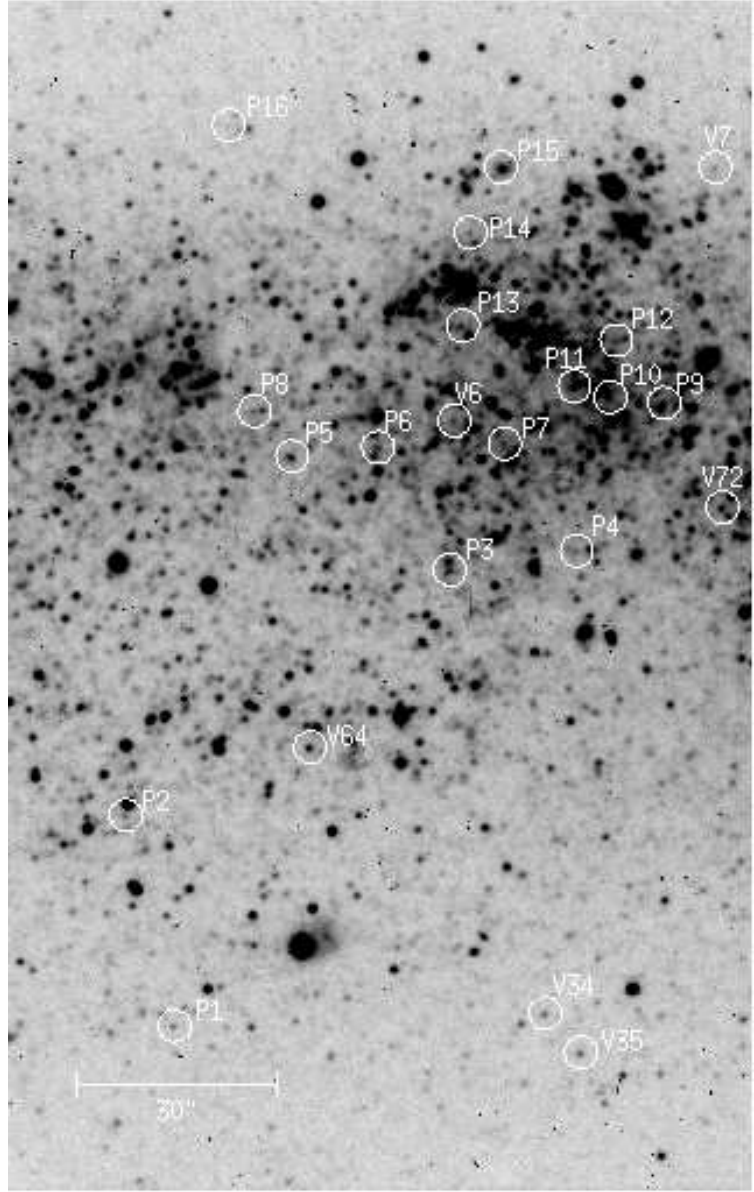} \caption{
$B$-band CCD image (2.2m telescope, 40 min exposure) of the field F1 in
NGC~3109. North is up, East to the right. The Cepheids are marked by a
circle. The stars coded as V belong to the original SC sample, while P
refers to the new candidate Cepheids.} \label{fig1} \end{figure}

\singlespace

\begin{deluxetable}{rcccc} \scriptsize \tablenum{1} \tablecolumns{5}
\tablewidth{0pc} \tablecaption{Log book of the observations.
\label{table1}} \tablehead{ \colhead{Date} & \colhead{Band} &
\colhead{Exp. time} & \colhead{Seeing} & \colhead{calib.}\\ \colhead{}  
  & \colhead{}     & \colhead{[min]}     & \colhead{FWHM} &
\colhead{err.\tablenotemark{A}}} \startdata  1-12-1991    
&I\tablenotemark{B}     &24   &$1'.14$&0.05\nl  1-12-1991     &R     &12
  &$1'.02$&0.05\nl  1-03-1992     &B     &40   &$1'.17$&0.03\nl 
1-03-1992     &V     &20   &$1'.16$&0.03\nl  2-03-1992     &B     &40  
&$1'.32$&0.03\nl  3-03-1992     &B     &40   &$1'.12$&0.03\nl  7-03-1992
    &B     &40   &$0'.87$&0.03\nl  8-03-1992     &B     &40  
&$1'.30$&0.03\nl  9-03-1992     &B     &40   &$0'.80$&0.03\nl  9-03-1992
    &R     &18   &$0'.94$&0.07\nl 17-02-1993     &B     &30  
&$1'.30$&0.03\nl 18-02-1993     &B     &30   &$1'.00$&0.03\nl 20-02-1993
    &B     &25   &$1'.60$&0.03\nl 20-02-1993     &I     &13  
&$1'.20$&0.05\nl \enddata \tablenotetext{A}{This is the total error on
the calibration (see text for details)} \tablenotetext{B}{This image is
a sum of two images each one with an exposition time of 12 min.}
\end{deluxetable}

The $I$ and $R$ band images of December 1991 were taken with EFOSC2 +
CCD \#17 at the ESO/MPI 2.2m telescope at La Silla in Chile. The CCD
format is $1024\times1024$ pixels of $0''.332$. The set of the $B$, $V$,
and $R$ band images of March 1992 was obtained with the RCA \#8 CCD
camera at the Cassegrain focus of the ESO/MPI 2.2m telescope with a CCD
format of 640$\times$1024 pixels and a pixel size of $0''.175$.
Furthermore, three $B$--band frames at the blue arm of EMMI equipped
with the Tektronix \#31 CCD and one in the $I$--band at the red arm of
EMMI with the Tektronik \#18 CCD have been collected at the ESO-NTT
telescope on February, 1993. CCD dimensions is $1024\times1024$ pixels
and the pixel sizes in blue and red are $0''.370$ and $0''.290$
respectively.

During these observing runs, at least thirteen Landolt's (1983a,b)
standard stars were observed in each color in order to transform the
instrumental photometry into the Landolt standard system.

A large set of bias, dark, and flat field frames have been collected,
particularly during the observing runs at the 2.2m telescope. Indeed,
the columns of the high resolution ESO-RCA chip \#8 have not a constant
bias as it depends on the total intensity of the light which falls over
the column. The procedure used for the image cleaning and calibration is
as in  PCP and Pellegrini (1993).

The CCD data have been reduced with DAOPHOT and ALLSTAR. Particular
attention has been devoted in tying the stellar photometry to Landolt's
(1983a,b) standard system. For the calibration of the $B$ and $V$ colors
we used the previous CPB calibration, while for the calibration of the
$R$-band and $I$-band we observed 16 Landolt (1983a,b) standard stars in
each color in December of 1991 and 13 Landolt standard stars for the $I$
band in February of 1993. As in Piotto et al. (1990), in order to obtain
the $R_{1991}$, $I_{1991}$ and $I_{1993}$ calibrated magnitudes we
adopted linear color terms. For the calibration of the $R_{1992}$
magnitudes we applied the same calibration parameters of $R_{1991}$,
after having obtained the linear relation between $R_{1991}$ and
$R_{1992}$. The uncertainty on the calibration of the $B$ and $V$
photometry is $0.03$~mag. This value represents the combination of the
error in the CPB calibration and of the error on the zero point
difference between our photometry and that of CPB. For the $R$ and $I$
bands we obtained a zero point error in the calibration equation of 0.01
mag. However, the largest error source in the calibration procedure
comes from the transformation of the zero point of the
point-spread-function (PSF) fitting photometry of NGC~3109 stars into
the standard star aperture photometry zero point (CPB). In our frames
all of the stars are distributed over a very inhomogeneous background,
making the aperture photometry quite uncertain. For the $I$ band we have
two independent calibrations (one from the 2.2m telescope and one from
the NTT telescope). The average zero point differences from night to
night were of 0.05 mag. The same (average) zero point calibration has
been adopted for the two $I$ frames. In $R$ we could not use the same
method, since the calibration equation was available for one night only.
For $R_{1991}$ we obtained an error of $0.05$~mag on the difference
between the aperture photometry and the PSF fitting photometry. Finally,
we obtained an uncertainty of $0.07$~mag on the calibration of the
$R_{1992}$ magnitude determinations summing the error on the linear
relation between $R_{1991}$ and $R_{1992}$ magnitudes to the calibration
error of the $R_{1991}$ calibration. Throughout the paper, whenever
needed, we took into proper account the fact that the zero point
calibration of $R_{1991}$ is more accurate then the $R_{1992}$ one. The
calibration zero point errors for each image are reported in
Table~\ref{table1}, Col.~5.

The internal error on the photometry can be easily evaluated by comparing
the independent magnitude determinations of the not variable stars in
each color. For each color, Table~\ref{table2}, gives the mean internal
errors (identified with {\it error}, Cols.~2,~4,~6~and~8) and the number
of stars (identified with {\it stars}, Cols.~3,~5,~7~and~9) at different
magnitude intervals (Col.~1) in the four photometric bands. The accuracy
of the CCD light curves of the Cepheids can be obtained from the
uncertainties quoted in this Table.

\begin{deluxetable}{ccrccrccrccr} \scriptsize \tablenum{2}
\tablecolumns{12} \tablewidth{0pc} \tablecaption{Photometric internal
errors in the B, V, R and I bands. \label{table2}} \tablehead{
\colhead{} & \multicolumn{2}{c}{B} &\colhead{} &\multicolumn{2}{c}{V}
&\colhead{} &\multicolumn{2}{c}{R} &\colhead{} &\multicolumn{2}{c}{I}\\
\cline{2-3} \cline{5-6} \cline{8-9} \cline{11-12}\\ \colhead{Mag} &
\colhead{Error} & \colhead{Stars} &\colhead{} & \colhead{Error} &
\colhead{Stars} &\colhead{}  & \colhead{Error} & \colhead{Stars}
&\colhead{}  & \colhead{Error} & \colhead{Stars}} \startdata   19.75  &
0.04  &  12 && 0.02 & 23&&0.02&42&&0.02&55\nl   20.25  & 0.04  &  28 &&
0.04 & 38&&0.03&66&&0.03&84\nl   20.75  & 0.04  &  83 && 0.04
&108&&0.04&140&&0.04&121\nl   21.25  & 0.04  & 118 && 0.05
&172&&0.05&150&&0.06&121\nl   21.75  & 0.05  & 157 && 0.06
&208&&0.06&163&&0.08&95\nl   22.25  & 0.06  & 189 && 0.07
&227&&0.10&76&&0.12&17\nl   22.75  & 0.07  & 192 && 0.09
&136&&0.20&18&&0.29&8\nl   23.25  & 0.08  & 134 && 0.14 &
50&&--&--&&--&--\nl   23.75  & 0.13  &  33 && 0.30 & 6&&--&--&&--&--\nl 
 24.25  & 0.17  &   8 && 0.32 & 1&&--&--&&--&--\nl \enddata
\end{deluxetable}

\section{Cepheids in NGC~3109}

\subsection{Cepheids in common with previous studies} \label{oldcefs} 
\begin{figure} \figurenum{2} \plotone{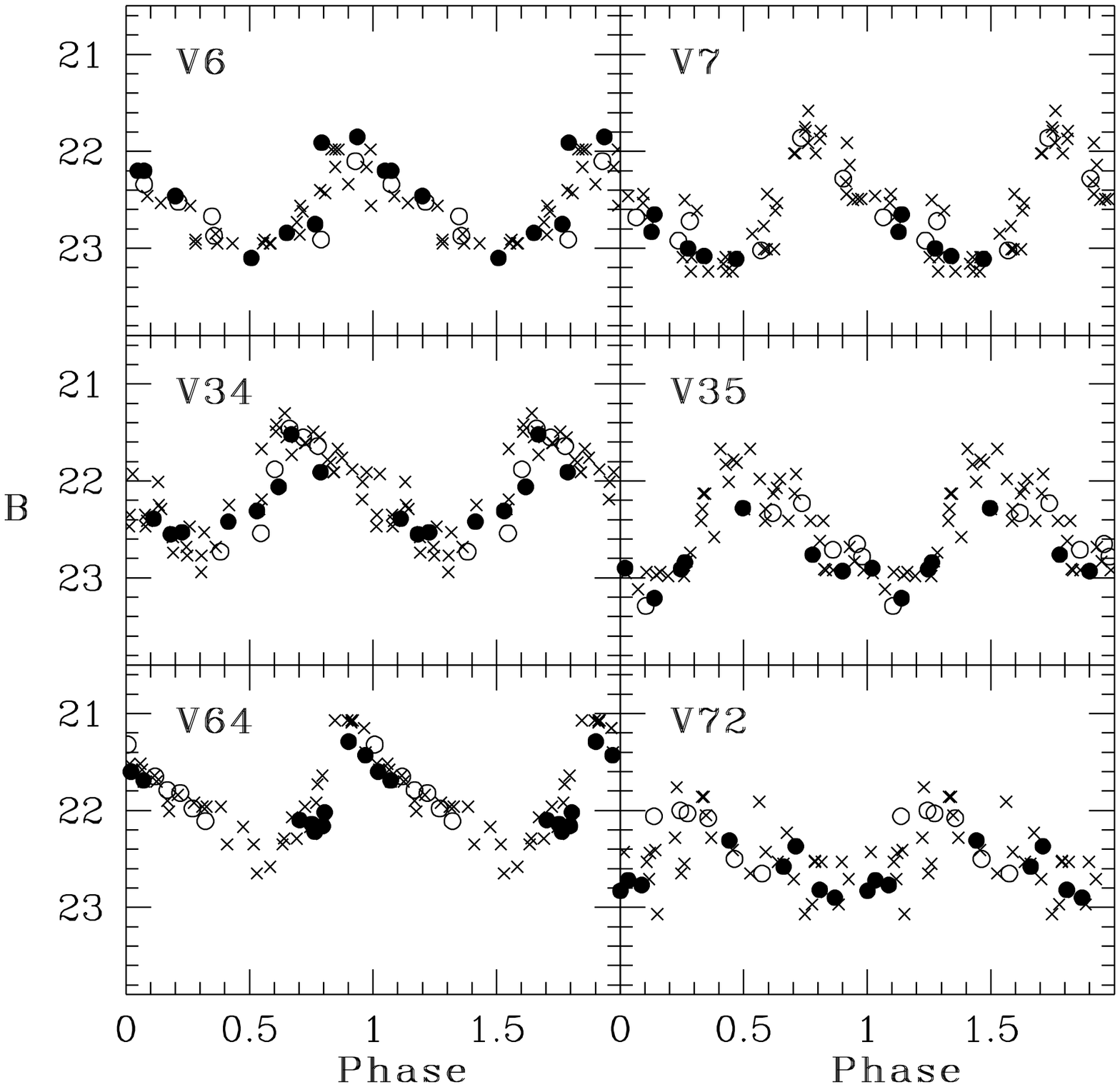} \caption {Light curves
for the six Cepheid variables identified by SC in field F1. {\it
Crosses} represent the original SC photometry, shifted by $\Delta
B=0.29$~mag (CPB). The {\it Open circles} represent the CPB data and the
{\it filled circles} the new data presented in this paper.} \label{fig2}
\end{figure}

SC
have identified 29 Cepheids in NGC~3109. CPB selected only 21 of them,
excluding candidate Cepheids with high photometric uncertainties or
those that did not show significant light variations during the period
covered by their observations (\cf\ CPB for more details). This work is
based on the same selection of 21 Cepheids. Six of them (marked as V in
Fig.~\ref{fig1}) fall in field F1 monitored in the present study. As in
CPB, we applied a zero point shift of $-0.29$ mag to the SC data (SC
data are fainter). The new CCD data in the $B$--band for the six
Cepheids located in the field F1 were plotted together with CPB and SC
photometry, and the periods were interactively adjusted in order to
obtain the best phase match among these three data sets. The new,
complete light curves of these six Cepheids are reproduced in
Fig.~\ref{fig2}. Differences between our and CPB periods are very small,
$\sim 10^{-2}$ days for V72 and $\sim 10^{-4}$ days for the other
variables (\cf\ Table~\ref{table5}).

For each of these variables, we could also add another point in the
$V$--band light curve, and one or two magnitude determinations in the
$I$ and $R$ bands. In addition, also SC V36 and V45 variables fall
within our larger $R$ and $I$ frames, allowing us to measure their $R$
and $I$ magnitudes. The new magnitude determinations for the SC Cepheids
are reported in the Table~\ref{table3}.

\begin{deluxetable}{rccccccccc} \scriptsize \tablenum{3}
\tablecolumns{10} \tablewidth{0pc} \tablecaption{SC Cepheids in NGC
3109\tablenotemark{A}. \label{table3}} \tablehead{ \colhead{Julian Date}
& \colhead{B} & \colhead{V} & \colhead{R} & \colhead{I} & \colhead{}
&\colhead{B} & \colhead{V} & \colhead{R} & \colhead{I} } \startdata
&\multicolumn{4}{c}{V6} & \colhead{} & \multicolumn{4}{c}{V7} \nl
\cline{1-10}\nl      2448591.825 & --   & --   & --   & --  & & --   &
--   &21.13  &20.94\nl \ph{244}8683.579 & --   &22.37  & --   & --  &
&22.65  &21.92  & --   &--\nl \ph{244}8684.779 &21.85  & --   & --   &
--  & &23.08  & --   & --   & --\nl \ph{244}8685.561 &22.20  & --   & --
  & --  & &23.11  & --   & --   & --\nl \ph{244}8688.789 &23.10  & --  
& --   & --  & & --   & --   & --   & --\nl \ph{244}8689.794 &22.84  &
--   & --   & --  & & --   & --   & --   & --\nl \ph{244}8690.786 &21.91
 & --   &21.11  & --  & & --   & --   &21.72  &--\nl \ph{244}9036.804
&22.20  & --   & --   & --  & &22.83  & --   & --   & --\nl
\ph{244}9037.688 &22.46  & --   & --   & --  & &23.00  & --   & --   &
--\nl \ph{244}9039.734 & --   & --   & --   &20.90 & & --   & --   & -- 
 & --  \nl \cline{1-10}\nl & \multicolumn{4}{c}{V34} &&
\multicolumn{4}{c}{V35}\nl \cline{1-10}\nl      2448591.825 & --   & -- 
 &20.54  &20.27 & & --   & --   &21.40  &21.20\nl \ph{244}8683.579
&22.39  & --   & --   & --  & &22.90  &22.07  & --   & -- \nl
\ph{244}8684.779 &22.55  & --   & --   & --  & & --   & --   & --   &
--\nl \ph{244}8685.561 &22.53  & --   & --   & --  & &22.84  & --   & --
  & --\nl \ph{244}8688.789 &22.42  & --   & --   & --  & & --   & --   &
--   & --\nl \ph{244}8689.794 & --   & --   & --   & --  & &22.76  & -- 
 & --   & --\nl \ph{244}8690.786 &22.31  & --   &20.42  & --  & &22.93 
& --   &22.66  &--\nl \ph{244}9036.804 &22.06  & --   & --   & --  &
&23.21  & --   & --   &--\nl \ph{244}9037.688 &21.52  & --   & --   & --
 & &22.91  & --   & --   &--\nl \ph{244}9039.734 &21.91  & --   & --  
&20.00 & &22.28  & --   & --   &21.08\nl \cline{1-10}\nl &
\multicolumn{4}{c}{V64}& & \multicolumn{4}{c}{V72}\nl \cline{1-10}\nl   
  2448591.825 & --   & --   &20.22  & 19.90 & & --   & --   &22.41 
&21.54\nl \ph{244}8683.579 &22.10  &21.18  & --   & --   & &22.90  & -- 
 & --   & --\nl \ph{244}8684.779 &22.22  & --   & --   & --   &&22.83  &
--   & --   & --\nl \ph{244}8685.561 &22.02  & --   & --   & --  
&&22.77  & --   & --   & --\nl \ph{244}8688.789 &21.43  & --   & --   &
--   &&22.31  & --   & --   & --\nl \ph{244}8689.794 &21.60  & --   & --
  & --   && --   & --   & --   & --\nl \ph{244}8690.786 &21.69  & --  
&20.23  & --   &&22.58  & --   & --   & --\nl \ph{244}9036.804 &22.14  &
--   & --   & --   &&22.37  & --   & --   & --\nl \ph{244}9037.688
&22.16  & --   & --   & --   &&22.82  & --   & --   & --\nl
\ph{244}9039.734 &21.29  & --   & --   &19.96  &&22.72  & --   & --   &
--  \nl \cline{1-10}\nl & \multicolumn{4}{c}{V36}& &
\multicolumn{4}{c}{V45}\nl \cline{1-10}\nl      2448591.825 & --   & -- 
 &22.41  &21.54  && --   & --   &22.50  & --   \nl \ph{244}8683.579 & --
  & --   & --   & --   && --   & --   & --   & --   \nl \ph{244}8684.779
& --   & --   & --   & --   && --   & --   & --   & --\nl
\ph{244}8685.561 & --   & --   & --   & --   && --   & --   & --   &
--\nl \ph{244}8688.789 & --   & --   & --   & --   && --   & --   & --  
& --\nl \ph{244}8689.794 & --   & --   & --   & --   && --   & --   & --
  & --\nl \ph{244}8690.786 & --   & --   & --   & --   && --   & --   &
--   & --\nl \ph{244}9036.804 & --   & --   & --   & --   && --   & --  
& --   & --\nl \ph{244}9037.688 & --   & --   & --   & --   && --   & --
  & --   & --\nl \ph{244}9039.734 & --   & --   & --   & --   && --   &
--   & --   &--\nl \enddata \tablenotetext{A}{For CPB data see their
Table 3(a).} \end{deluxetable}

\subsection{New candidate Cepheids} \label{newcefs} In order to search
for new candidate Cepheid variables, we used both CPB raw data and the
photometry presented in this paper. We computed for each star, for both
the $B$ and $V$ magnitude determinations, the {\it normalized} standard
deviation ($\sigma_N$), i.e. the standard deviation divided by the mean
photometric internal error in each magnitude intervals of 0.5 mag. We
considered as possible variables only those stars satisfying the
relation: $$\sqrt{\sigma_N^2(B)+\sigma_N^2(V)} \ge 2$$

\begin{figure} \figurenum{3} \plotone{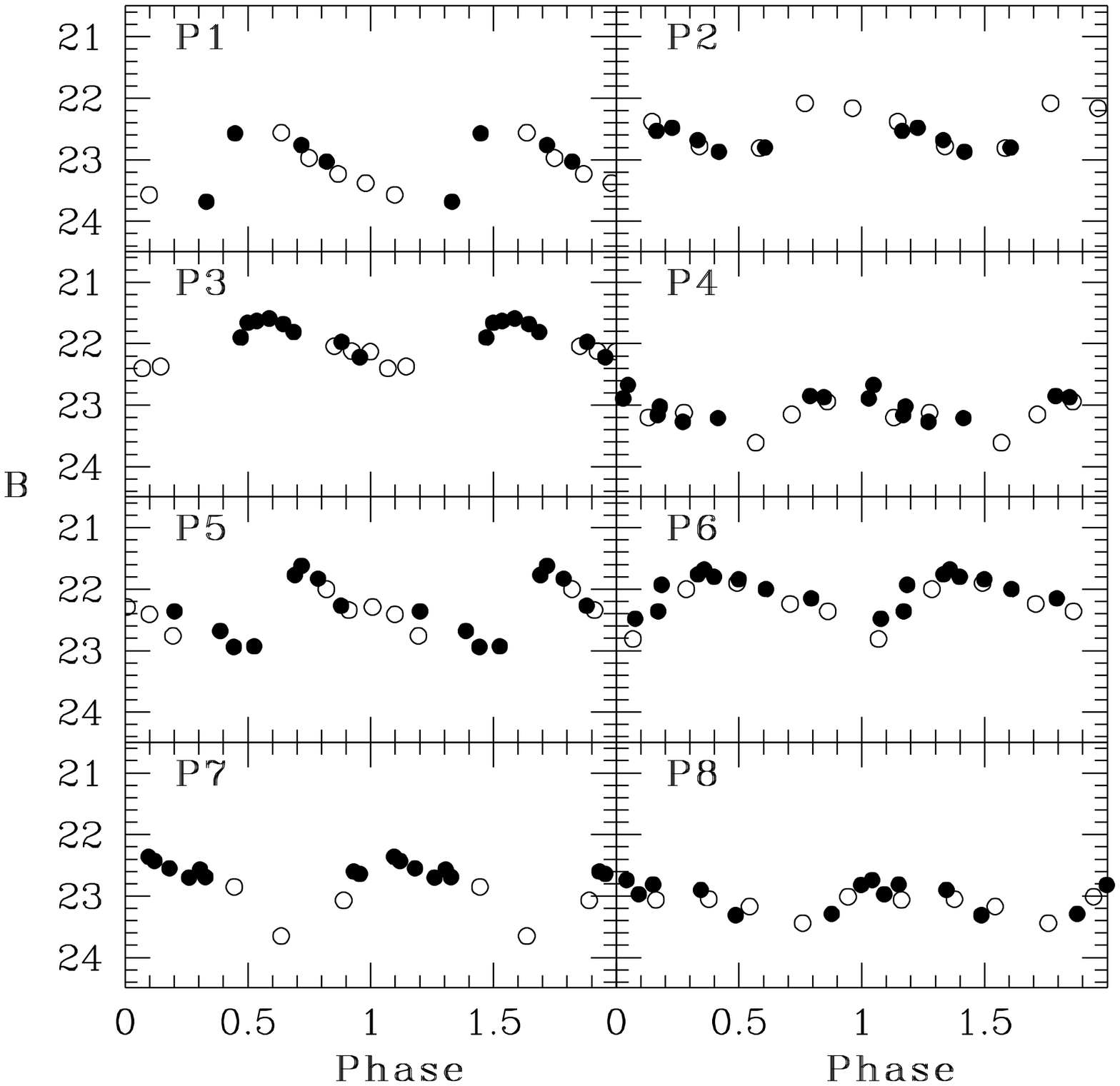} \begin{center} {\bf a)}
\end{center} \plotone{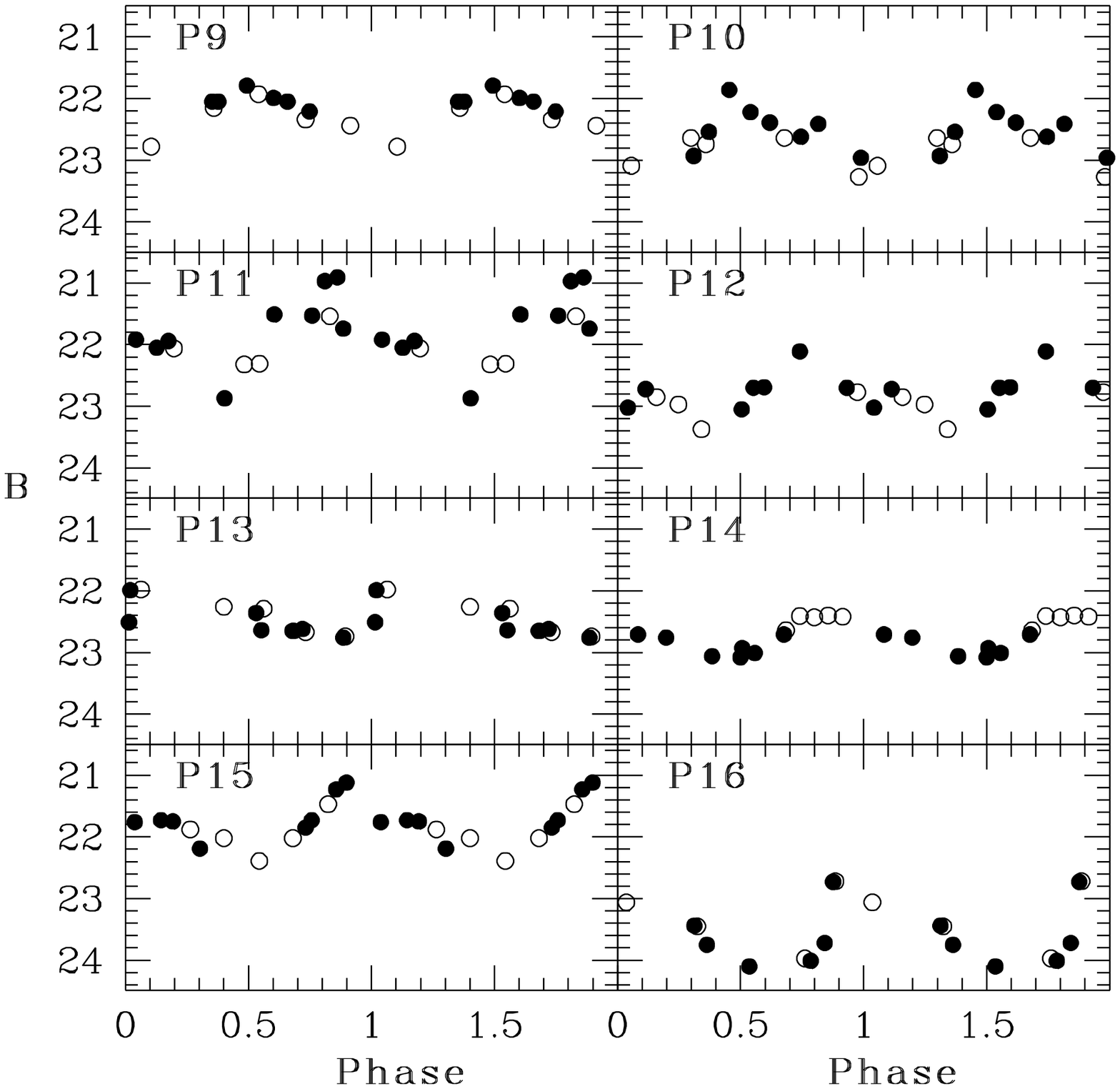} \begin{center} {\bf b)} \end{center}
\caption {a), b): The Light curves for the 16 new candidate Cepheid
variables with a tentative period identification ({\it cf.} Table 5).
The {\it open circles} represent the CPB's data, while the {\it filled
circles} are for the data presented in this paper.} \label{fig3}
\end{figure}

By this procedure, we isolated 129 stars. The possible Cepheids were
selected from these candidates on the basis of their light curve,
excluding spurious objects in which only one or two points contributed
to the large deviation. We ended with a list of 29 stars. Among these
there were all the Cepheids previously discovered by SC. Of the
remaining variable candidates, only 16 had colors and light curves
(after rephasing) typical of a Cepheid\footnote{$(B-V)$ colors have
proven to be a valuable aid in discriminating Cepheids from other
variables, allowing the rejection of many objects with colors differing
significantly from those typical of Cepheids (\cf\  also
Fig.~\ref{fig11}).}. The 16 new Cepheid candidates located in field F1
are marked with a circle and coded as P in Fig.~\ref{fig1}; their
photometry is reported in Table~\ref{table4}. The most probable periods
of the new Cepheids have been determined by the usual Fourier analysis. 
More data are needed to confirm the periods, though the light curves
seem well enough covered to allow a fair determination of the parameters
of the PL relation.  The light curves of these new candidates are
plotted with their tentative periods in Fig.~\ref{fig3}.

\begin{deluxetable}{rccccccccc} \scriptsize \tablenum{4}
\tablecolumns{10} \tablewidth{0pc} \tablecaption{New candidate Cepheids
in NGC 3109: CPB and our data. \label{table4}} \tablehead{
\colhead{Julian Date} & \colhead{B} & \colhead{V} & \colhead{R} &
\colhead{I} & \colhead{} & \colhead{B} & \colhead{V} & \colhead{R} &
\colhead{I} } \startdata & \multicolumn{4}{c}{P1} & \colhead{} &
\multicolumn{4}{c}{P2} \nl \cline{1-10}\nl      2447967.588 & 22.56   &
22.02   & --   & --   && 22.81   & 22.40   & --   & --\nl
\ph{244}7968.559 & 22.97   & 22.14   & --   & --   && 22.08   & 21.88  
& --   & --\nl \ph{244}7969.581 & 23.23   & 22.44   & --   & --   &&
22.16   & 21.98   & --   & --\nl \ph{244}7970.551 & 23.38   & 22.49   &
--   & --   && 22.38   & 22.02   & --   & --\nl \ph{244}7971.570 & 23.57
  & 22.63   & --   & --   && 22.78   & 22.36   & --   & --\nl
\ph{244}8591.825 & --   & --   &22.30  &22.07  && --   & --   & 22.01  
& --\nl \ph{244}8683.579 & --   & --   & --   & --   && --   & --   & --
  & --\nl \ph{244}8684.779 & --   & --   & --   & --   && --   & --   &
--   & --\nl \ph{244}8685.561 & --   & --   & --   & --   && --   & --  
& --   & --\nl \ph{244}8688.789 &23.68  & --   & --   & --   &&22.48  &
--   & --   & --\nl \ph{244}8689.794 &22.57  & --   & --   & --  
&&22.87  & --   & --   & --\nl \ph{244}8690.786 & --   & --   & --   &
--   &&22.80  & --   & --   & --\nl \ph{244}9036.804 &22.76  & --   & --
  & --   &&22.53  & --   & --   & --\nl \ph{244}9037.688 &23.03  & --  
& --   & --   &&22.68  & --   & --   & --\nl \ph{244}9039.734 & --   &
--   & --   &21.90  && --   & --   & --   & 21.83 \nl \cline{1-10}\nl &
\multicolumn{4}{c}{P3} && \multicolumn{4}{c}{P4}\nl \cline{1-10}\nl     
2447967.588 & 22.04   & 21.27   & --   & --   && 22.94   & 23.16   & -- 
 & --\nl \ph{244}7968.559 & 22.12   & 21.28   & --   & --   && 23.12   &
--   & --   & --\nl \ph{244}7969.581 & 22.13   & --   & --   & --   &&
23.15   & --   & --   & --\nl \ph{244}7970.551 & 22.40   & 21.40   & -- 
 & --   && 23.20   & 22.83   & -- & --\nl \ph{244}7971.570 & 22.37   &
21.55   & --   & --   && 23.61   & 23.03   & --   & --\nl
\ph{244}8591.825 & --   & --   &20.65  &20.42  && --   & --   & --   &
--\nl \ph{244}8683.579 &21.66  &20.97  & --   & --   &&23.02  & --   &
--   & -- \nl \ph{244}8684.779 &21.59  & --   & --   & --   && --   & --
  & --   & --\nl \ph{244}8685.561 &21.68  & --   & --   & --   &&22.89 
& --   & --   & -- \nl \ph{244}8688.789 &21.97  & --   & --   & --  
&&23.21  & --   & --   & --\nl \ph{244}8689.794 &22.22  & --   & --   &
--   &&22.87  & --   & --   & --\nl \ph{244}8690.786 & --   & --  
&20.94  & --   &&23.27  & --   & --   & --\nl \ph{244}9036.804 &21.90  &
--   & --   & --   &&22.85  & --   & --   & --\nl \ph{244}9037.688
&21.63  & --   & --   & --   &&23.16  & --   & --   & --\nl
\ph{244}9039.734 &21.81  & --   & --   &20.36  &&22.67  & --   & --   &
--  \nl \cline{1-10}\nl & \multicolumn{4}{c}{P5} &&
\multicolumn{4}{c}{P6}\nl \cline{1-10}\nl      2447967.588 & 22.00   &
21.29   & --   & --   && 22.36   & 21.91   & --   & --\nl
\ph{244}7968.559 & 22.34   & 21.53   & --   & --   && 22.81   & 21.95  
& --   & --\nl \ph{244}7969.581 & 22.29   & 21.51   & --   & --   &&
22.00   & 21.51   & --   & --\nl \ph{244}7970.551 & 22.41   & --   & -- 
 & --   && 21.90   & 21.70   & --   & --\nl \ph{244}7971.570 & 22.76   &
21.96   & --   & --   && 22.24   & 21.69   & --   & --\nl
\ph{244}8591.825 & --   & --   & --   &20.74  && --   & --   &21.36 
&21.00\nl \ph{244}8683.579 &22.36  &21.55  & --   & --   &&22.48  & --  
& --   & --\nl \ph{244}8684.779 & --   & --   & --   & --   &&21.76  &
--   & --   & --\nl \ph{244}8685.561 &22.68  & --   & --   & --  
&&21.84  & --   & --   & --\nl \ph{244}8688.789 &21.77  & --   & --   &
--   &&21.93  & --   & --   & --\nl \ph{244}8689.794 &21.83  & --   & --
  & --   &&21.80  & --   & --   & --\nl \ph{244}8690.786 &22.27  & --  
&21.07  & --   &&22.00  & --   &21.37  & --\nl \ph{244}9036.804 &22.94 
& --   & --   & --   &&22.36  & --   & --   & --\nl \ph{244}9037.688
&22.93  & --   & --   & --   &&21.68  & --   & --   & --\nl
\ph{244}9039.734 &21.62  & --   & --   &20.64  &&22.15  & --  & --  
&21.04 \nl \cline{1-10}\nl \tablebreak & \multicolumn{4}{c}{P7} &&
\multicolumn{4}{c}{P8}\nl \cline{1-10}\nl      2447967.588 & 23.07   &
21.71   & --   & --   && 23.05   & 22.64   & --   & --\nl
\ph{244}7968.559 & --   & --   & --   & --         && 23.44   & --   &
--   & --\nl \ph{244}7969.581 & --   & 21.62   & --   & --      && 23.06
  & 22.84   & --   & --\nl \ph{244}7970.551 & 22.85   & 21.64   & --   &
--   && 23.17   & --   & --   & --\nl \ph{244}7971.570 & 23.65   & 21.59
  & --   & --   && 23.01   & 23.02   & --   & --\nl \ph{244}8683.579
&22.64  & 21.48   & --   & --   &&22.74  & --   & --   & --\nl
\ph{244}8684.779 &22.55  & --   & --   & --   && --   & --   & --   &
--\nl \ph{244}8685.561 &22.69  & --   & --   & --   && --   & --   & -- 
 & --\nl \ph{244}8688.789 &22.60  & --   & --   & --   &&22.97  & --   &
--   & --\nl \ph{244}8689.794 &22.43  & --   & --   & --   &&23.31  & --
  & --   & --\nl \ph{244}8690.786 &22.57  & --   & --  & --   &&23.29  &
--   &--& --\nl \ph{244}9036.804 &22.36  & --   & --   & --   &&22.82  &
--   & --   & --\nl \ph{244}9037.688 &22.70  & --   & --   & --  
&&22.90  & --   & --   & --\nl \ph{244}9039.734 & --   & --   & --   &
--   &&22.81  & --   & --   & --  \nl \cline{1-10}\nl \colhead{} &
\multicolumn{4}{c}{P9} & \colhead{} & \multicolumn{4}{c}{P10} \nl
\cline{1-10}\nl      2447967.588 & 22.15   & 21.69   & --   & --   &&
23.09   & --   & --   & --\nl \ph{244}7968.559 & 21.93   & --   & --   &
--      && 22.74   & --   & --   & --\nl \ph{244}7969.581 & 22.34   & --
  & --   & --      && 22.64   & 21.69   & --   & --\nl \ph{244}7970.551
& 22.44   & 21.76   & --   & --   && 23.27   & 21.91   & --   & --\nl
\ph{244}7971.570 & 22.78   & 21.80   & --   & --   && 22.64   & 21.66  
& --   & --\nl \ph{244}8591.825 & --   & --   &21.00  & --  && --   & --
  & --   &20.48\\ \ph{244}8683.579 &22.05  &21.50  & --   & --   &&22.54
 &21.61  & --   & --\\ \ph{244}8684.779 &21.99  & --   & --   & --  
&&22.62  & --   & --   & --\\ \ph{244}8685.561 &22.21  & --   & --   &
--   &&22.96  & --   & --   & --\\ \ph{244}8688.789 &22.05  & --   & -- 
 & --   && --   & --   & --   & -- \\ \ph{244}8689.794 & --   & --   &
--   & --   &&22.93  & --   & --   & --\\ \ph{244}8690.786 & --   & --  
&20.99  & --   &&22.39  & --   &21.29  & --\\ \ph{244}9036.804 &21.79  &
--   & --   & --   &&22.22  & --   & --   & --\\ \ph{244}9037.688 &22.05
 & --   & --   & --   &&22.41  & --   & --   & --\\ \ph{244}9039.734 &
--   & --   & --   &20.77  &&21.86  & --   & --   &20.35\\
\cline{1-10}\nl & \multicolumn{4}{c}{P11}& & \multicolumn{4}{c}{P12}\nl
\cline{1-10}\nl      2447967.588 & 22.32   & 21.59   & --   & --   &&
22.77   & --   & --   & --\nl \ph{244}7968.559 & 21.54   & 21.01   & -- 
 & --   && --      & --   & --   & --\nl \ph{244}7969.581 & 22.06   &
22.12   & --   & --   && 22.85   & 22.74   & --   & --\nl
\ph{244}7970.551 & 22.31   & 21.78   & --   & --   && 22.97   & 22.98  
& --   & --\nl \ph{244}7971.570 & --   & 21.59   & --   & --      &&
23.37   & 22.92   & --   & --\nl \ph{244}8591.825 & --   & --   & --   &
--  & & --   & --   &22.49  & --\nl \ph{244}8683.579 &21.94  &21.73  &
--   & --  & &22.70  &22.32  & --   & --\nl \ph{244}8684.779 &21.51  &
--   & --   & --  & &23.02  & --   & --   & --\nl \ph{244}8685.561
&21.74  & --   & --   & --  & &22.72  & --   & --   & --\nl
\ph{244}8688.789 &21.92  & --   & --   & --  & & --   & --   & --   &
--\nl \ph{244}8689.794 &22.87  & --   & --   & --  & &23.05  & --   & --
  & --\nl \ph{244}8690.786 &21.53  & --   &21.38  & --  & &22.69  & --  
&22.24  & --  \nl \ph{244}9036.804 &20.97  & --   & --   & --  & & --  
& --   & --   & --\nl \ph{244}9037.688 &22.05  & --   & --   & --  &
&22.70  & --   & --   & --\nl \ph{244}9039.734 &20.91  & --   & --   &
--  & &22.11  & --   & --   & --\nl \cline{1-10}\nl \tablebreak &
\multicolumn{4}{c}{P13}& & \multicolumn{4}{c}{P14}\nl \cline{1-10}\nl   
  2447967.588 & 22.26   & 21.80   & --   & --   && 22.64   & 22.37   &
--   & --\nl \ph{244}7968.559 & 22.29   & 21.57   & --   & --   && 22.41
  & 22.24   & --   & --\nl \ph{244}7969.581 & 22.67   & 21.78   & --   &
--   && 22.43   & 22.08   & --   & --\nl \ph{244}7970.551 & 22.74   &
22.22   & --   & --   && 22.40   & 22.04   & --   & --\nl
\ph{244}7971.570 & 21.98   & 21.63   & --   & --   && 22.42   & 21.87  
& --   & --\nl \ph{244}8591.825 & --   & --   &21.35  &21.11 & & --   &
--   &22.20  & --\nl \ph{244}8683.579 &22.65  & --   & --   & --  &
&22.71  & --   & --   & --\nl \ph{244}8684.779 & --   & --   & --   & --
 & & --   & --   & --   & --\nl \ph{244}8685.561 &22.51  & --   & --   &
--  & &22.76  & --   & --   & --\nl \ph{244}8688.789 &22.64  & --   & --
  & --  & &23.06  & --   & --   & --\nl \ph{244}8689.794 &22.62  & --  
& --   & --  & & --   & --   & --   & --\nl \ph{244}8690.786 &22.76  &
--   &21.65  & --  & &23.08  & --   &22.41  & --\nl \ph{244}9036.804
&22.36  & --   & --   & --  & &22.93  & --   & --   & --\nl
\ph{244}9037.688 &22.65  & --   & --   & --  & &23.01  & --   & --   &
--\nl \ph{244}9039.734 &21.99  & --   & --   & --  & &22.71  & --   & --
  &20.55\nl \cline{1-10}\nl & \multicolumn{4}{c}{P15}& &
\multicolumn{4}{c}{P16}\nl \cline{1-10}\nl      2447967.588 & 21.88   &
21.15   & --   & --   && 23.97   & 23.28   & --   & --\nl
\ph{244}7968.559 & 22.02   & 21.44   & --   & --   && 23.06   & 22.69  
& --   & --\nl \ph{244}7969.581 & 22.39   & --      & --   & --   &&
23.45   & 22.85   & --   & --\nl \ph{244}7970.551 & 22.02   & 21.42   &
--   & --   && --      & 23.20   & --   & --\nl \ph{244}7971.570 & 21.47
  & 21.09   & --   & --   && 22.72   & 22.33   & --   & --\nl
\ph{244}8591.825 & --   & --   &21.16  & --  & & --   & --   &22.51  &
--\nl \ph{244}8683.579 & --   & --   & --   & --  & &23.72  &23.36  & --
  & --\nl \ph{244}8684.779 &21.75  & --   & --   & --  & & --   & --   &
--   & --\nl \ph{244}8685.561 &22.19  & --   & --   & --  & & --   & -- 
 & --   & --\nl \ph{244}8688.789 &21.73  & --   & --   & --  & &23.44  &
--   & --   & --\nl \ph{244}8689.794 &21.12  & --   & --   & --  & & -- 
 & --   & --   & --\nl \ph{244}8690.786 &21.76  & --   & --   & --  &
&22.73  & --   & --   & --\nl \ph{244}9036.804 &21.85  & --   & --   &
--  & &24.10  & --   & --   & --\nl \ph{244}9037.688 &21.23  & --   & --
  & --  & &24.01  & --   & --   & --\nl \ph{244}9039.734 &21.73  & --  
& --   & --  & &23.75  & --   & --   &22.18 \nl \enddata
\end{deluxetable}

\section{Cepheid parameters} \label{parcef} The old photographic and the
new CCD $B$--band data provide a good coverage of the blue light curves.
It is therefore straightforward to derive the magnitudes at maximum
($B_{max}$) and the mean magnitudes ($<B>$) listed in Cols.~4 and 5 of
Table~\ref{table5}. The parameters for the SC Cepheids have been
estimated giving more weight to the CCD than to the photographic
photometry, particularly at the minimum light. We also attempted to
obtain the phase weighted mean magnitude using the relation (4) in Saha
et al. (1994) after transform the magnitudes into intensities. As
discussed by Saha \& Hoessel (1990), this method takes care of the bias
introduced by the loss of faint measurements. Unfortunately, the phase
sampling of the light curves is not sufficiently even; thus, we
preferred to compute the mean directly from the light curve drawn by
hand trough the available points and use the phase weighted mean only
for a comparison. In order to obtain the mean $V$--band magnitude we
used the same method as for the $B$--band. These values are reported in
Col.~7 of Table~\ref{table5}.

\begin{deluxetable}{rccccccccccc} \scriptsize \tablenum{5}
\tablecolumns{12} \tablewidth{0pc} \tablecaption{NGC~3109 Cepheid
parameters. \label{table5}} \tablehead{ \colhead{Cepheid}  &
\colhead{$P_{CPB}$} & \colhead{$P_{our}$}& \colhead{$B_{max}$}&
\colhead{$<B>$}&\colhead{$\sigma_{<B>}$}&
\colhead{$<V>$}&\colhead{$\sigma_{<V>}$}& \colhead{$<R>$}&
\colhead{$\sigma_{<R>}$\tablenotemark{A}} &
\colhead{$<I>$}&\colhead{$\sigma_{<I>}$\tablenotemark{A}}\\  
\colhead{Ident}   & \colhead{[days]}  & \colhead{[days]}
&\colhead{[mag]}&
\colhead{[mag]}&\colhead{[mag]}&\colhead{[mag]}&\colhead{[mag]}&\colhead{[mag]}
&\colhead{[mag]}&\colhead{[mag]}&\colhead{[mag]} } \startdata    V6  &
7.0223& 7.02120 &21.78 &22.46& 0.04&21.80& 0.04 &20.92& 0.29 &20.77&
0.23\nl    V7  & 5.9879& 5.98810 &21.71 &22.43& 0.05&21.98& 0.05 &21.22&
0.32 &20.91& 0.24\nl    V8  &14.4500&--  &20.85 &21.45& 0.10&--&--&--&--
\nl    V9  & 7.7670&--  &22.10 &22.90& 0.10&21.90& 0.10 &--&--&-- \nl  
V11  & 7.9298&--  &21.85 &22.40& 0.10&--&--&--&-- \nl   V12  & 8.1183&--
 &21.70 &22.43& 0.10&21.86& 0.10 &--&--&-- \nl   V18  & 8.3707&-- 
&21.60 &22.38& 0.10&21.68& 0.10 &--&--&-- \nl   V20  & 8.2718&--  &21.90
&22.40& 0.10&22.20& 0.10 &--&--&-- \nl   V34  &17.2320&17.22550 &21.43
&22.12& 0.04&21.28& 0.05 &20.62& 0.30 &20.12& 0.23\nl   V35  & 8.1928&
8.19190 &21.65 &22.45& 0.05&21.80& 0.05 &21.53& 0.35 &21.22& 0.27\nl  
V36  & 7.1290&--  &21.90 &22.25& 0.10&21.82& 0.10 &21.62& 0.15 &21.40&
0.12\nl   V44  & 9.0970&--  &21.10 &21.80& 0.10&21.27& 0.10 &--&--&-- \nl
  V45  & 8.7620&--  &21.50 &22.05& 0.10&21.36& 0.10 &21.04& 0.19 &--&--
\nl   V57  & 7.2595&--  &21.35 &21.70& 0.10&--&--&--&-- \nl   V64 
&19.5745&19.57070 &21.05 &21.79& 0.04&20.93& 0.04 &20.44& 0.32 &20.02&
0.25\nl   V72  & 8.6805& 9.09400 &21.72 &22.32& 0.04&21.76& 0.04 &21.34&
0.26 &21.02& 0.20\nl   V77  & 5.8170&--  &22.10 &22.60& 0.10&22.00& 0.10
&--&--&-- \nl   V79  & 8.2480&--  &21.95 &22.55& 0.10&21.70& 0.10
&--&--&-- \nl   V81  &19.9570&--  &21.15 &21.88& 0.10&21.12& 0.10
&--&--&-- \nl   V92  &13.3165&--  &21.45 &21.93& 0.10&21.15& 0.10
&--&--&-- \nl    P1  &--& 8.61700 &22.30 &22.99& 0.05&22.28& 0.05
&22.07& 0.30 &21.76& 0.23\nl    P2  &--& 5.27800 &22.00 &22.50&
0.04&22.18& 0.04 &22.10& 0.22 &21.87& 0.17\nl    P3  &--&13.60000 &21.50
&22.00& 0.03&21.24& 0.03 &20.77& 0.22 &20.36& 0.17\nl    P4  &--&
2.32980 &22.55 &23.13& 0.04&23.06& 0.06 &--&--&-- \nl    P5 
&--&10.62600 &21.60 &22.30& 0.04&21.64& 0.04 &20.91& 0.30 &20.30& 0.24\nl
   P6  &--& 4.70380 &21.65 &22.23& 0.03&21.68& 0.04 &21.40& 0.25 &21.00&
0.20\nl    P7  &--& 5.34060 &22.31 &23.00& 0.04&21.65& 0.04 &--&--&-- \nl
   P8  &--&2.54200  &22.71 &23.07& 0.03&22.79& 0.04 &--&--&-- \nl    P9 
&--& 5.34250 &21.74 &22.42& 0.04&21.73& 0.04 &20.93& 0.30 &20.61& 0.23\nl
  P10  &--& 3.20620 &21.85 &22.63& 0.04&21.68& 0.05 &21.13& 0.34 &20.53&
0.26\nl   P11  &--& 2.78930 &21.10 &22.00& 0.05&21.58& 0.06 &21.12& 0.40
&--&-- \nl   P12  &--&10.85540 &21.80 &22.57& 0.04&22.27& 0.05 &21.86&
0.34 &--&-- \nl   P13  &--& 6.00240 &21.89 &22.33& 0.03&21.81& 0.04
&21.26& 0.19 &21.00& 0.15\nl   P14  &--&17.29510 &22.20 &22.65&
0.03&--&-- &22.09& 0.20 &--&-- \nl   P15  &--& 7.10590 &21.10 &21.80&
0.04&21.30& 0.04 &21.19& 0.31 &--&-- \nl   P16  &--& 3.54310 &22.70
&23.41& 0.06&22.89& 0.06 &22.32& 0.31 &21.85& 0.24\nl \enddata
\tablenotetext{A}{This error is an upper limit, in fact it is the
half-amplitude of the light curve in this band, obtained using the
information 
from the blue light curves (see Sect. 3 for details).} 
 
\end{deluxetable}

Finally, even though there are only random phase data available in $R$
and $I$ bands, it is always possible to use the information from the blue
light curves to correct these observations to the mean $<R>$, $<I>$ light
(Freedman 1988, F88). The Cepheid light curves in blue differ from those
at longer wavelength in two respects: {\it (i)} the amplitude decreases
with increasing wavelength, and {\it (ii)} there is a phase shift moving
from the $B$ to the $I$--band. After transforming the blue photographic
light curves into the CCD photometric scale, a correction for both the
amplitude scale and the phase shift\footnote{We have assumed that the
amplitude ratios as a function of the color and the phase shift are the
same as in Freedman (1988).} was applied as in F88 in order to obtain
the mean magnitudes for the $R$ and $I$ data (Cols. 9 and 11 of
Table~\ref{table5}). Here, we took into proper account that the
$R_{1991}$ magnitude determinations are more accurate then the
$R_{1992}$ ones.

In order to estimate the uncertainty associated with the mean
magnitudes, we considered all the independent sources of errors acting
on this quantity: the amplitude of the light curve that was converted
into its equivalent variance (see Freedman et al. 1991, 1992), the
internal photometric errors, and the zero-point errors in the calibrated
magnitudes (see Table~\ref{table1} and Table~\ref{table2}). For the
bands $B$ and $V$, the total error was obtained adding in quadrature
these errors and dividing by the reduced number of degrees of freedom.
For the bands $R$ and $I$, we have so few points (no more than two) that
we consider more reasonable assuming as error on the mean magnitudes the
half-amplitude of the light curve, as predicted from the $B$ light curve
(F88). Obviously, this uncertainty must be considered as an upper limit
of the error to be associated to the mean magnitude. The resulting
errors are shown in Table 5 Cols. 6,8,10, and 12.

\section{The distance modulus}

\subsection{The relative apparent distance moduli} \label{relmod}
Recently, it has been repeatedly pointed out that the principal source
of systematic errors in the calibration of the primary distance
indicators is in the uncertainty in determination of the zero points of
the Cepheid Period-Luminosity and Period-Luminosity-Color (PLC)
relations (Freedman \& Madore 1993, Fukugita et al. 1993). For this
reason we prefer to derive the distance modulus of NGC~3109 by
comparison with the Large Magellanic Cloud Cepheids, as we did in the
other papers of this series (CPB, PCP) and as it was originally
suggested by F88. Indeed the distance modulus of the LMC can be derived
using several independent methods (and owing to the large number of
known Cepheids in the Magellanic Clouds, these galaxies represent almost
ideal laboratories for the calibrations of PL and PLC relations). The
method has the advantage of leaving the relative distance unaffected by
whatever improvement there will be in the distance of the LMC. According
to Hunter \& Gallagher (1985), the metal content of NGC~3109 should be
just $\sim 1.5$ times higher than in the LMC. Therefore, we do not
expect significant effects on the PL relations for the two galaxies. In
fact, Freedman \& Madore (1990) have shown that the distance moduli
obtained via Cepheid PL relation in a study of three fields of M31 with
metal content from 0.5 to 2.5 times the solar metallicity range over
less than 0.27 mag.

\begin{figure} \figurenum{4} \plotone{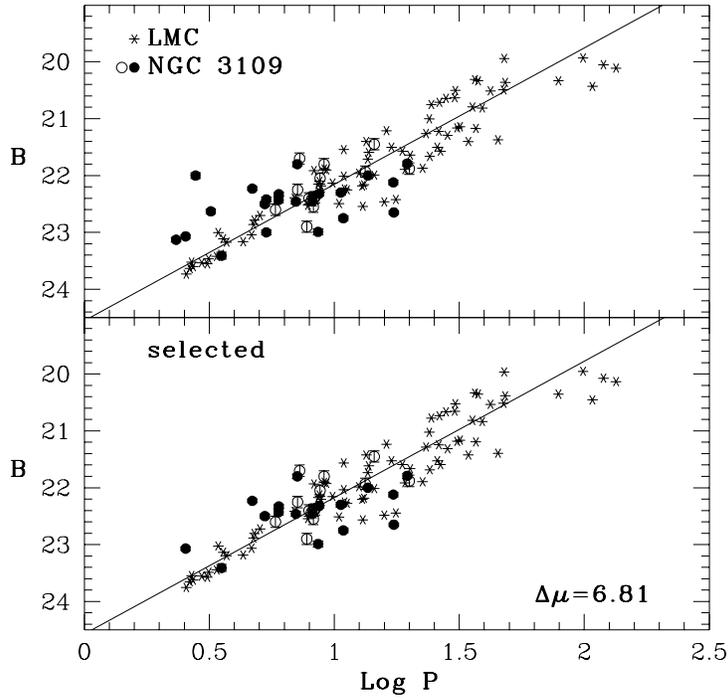} \caption {The
Period-Luminosity relation in the $B$-band for the NGC~3109 Cepheids is
compared with the corresponding relation for the LMC (from Sandage 1988
and Madore \& Freedman, 1991). {\it Filled circles} represent both SC
and our Cepheids located in the field F1 of NGC~3109, {\it open circles}
the SC Cepheids located in the other fields. The magnitude error bars
have dimensions of the order of the symbol size. {\it Crosses} reproduce
the photoelectric data for the LMC Cepheids shifted by the zero point
difference between the NGC~3109 and the LMC PL relations. In the {\it
upper panel} all the variables are used, while in the {\it lower panel}
only the Cepheids with the most reliable light curve parameters are
plotted. A best fit (see text) of the selected sample to the LMC
Cepheids with $\log P < 1.8$ gives us an apparent relative distance
modulus $\Delta\mu_B=6.81\pm 0.11$.} \label{fig4} \end{figure}

\begin{figure} \figurenum{5} \plotone{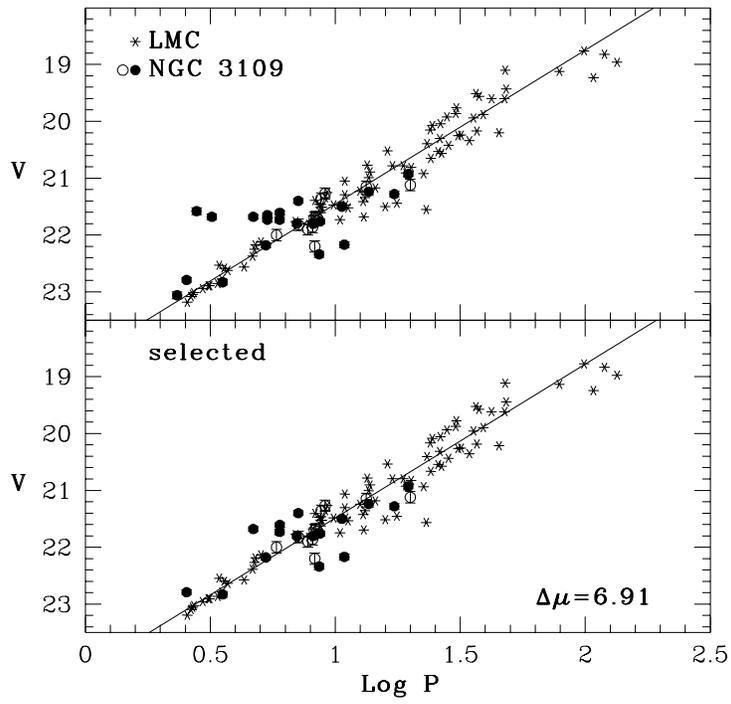} \caption{As in Fig. 4,
but for the $V$-band. The resulting apparent relative distance modulus
is $\Delta\mu_V=6.91 \pm 0.09$.} \label{fig5} \end{figure}

\begin{figure} \figurenum{6} \plotone{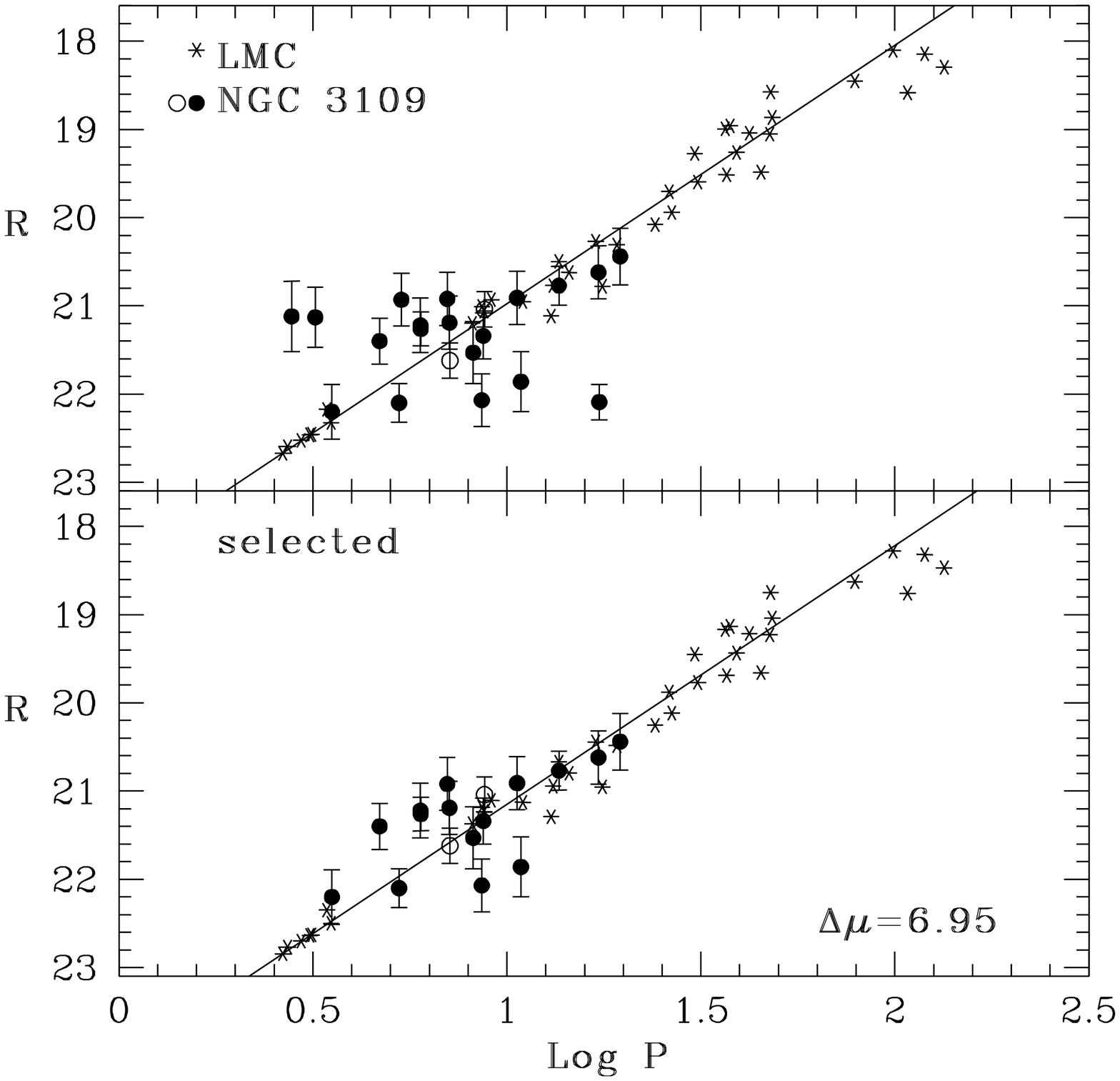} \caption{As in Fig. 4,
but for the $R$-band (the plotted errors are an upper limit, see Sect. 4
for details). The resulting apparent relative distance modulus is
$\Delta\mu_R=6.95 \pm 0.13$.} \label{fig6} \end{figure}

\begin{figure} \figurenum{7} \plotone{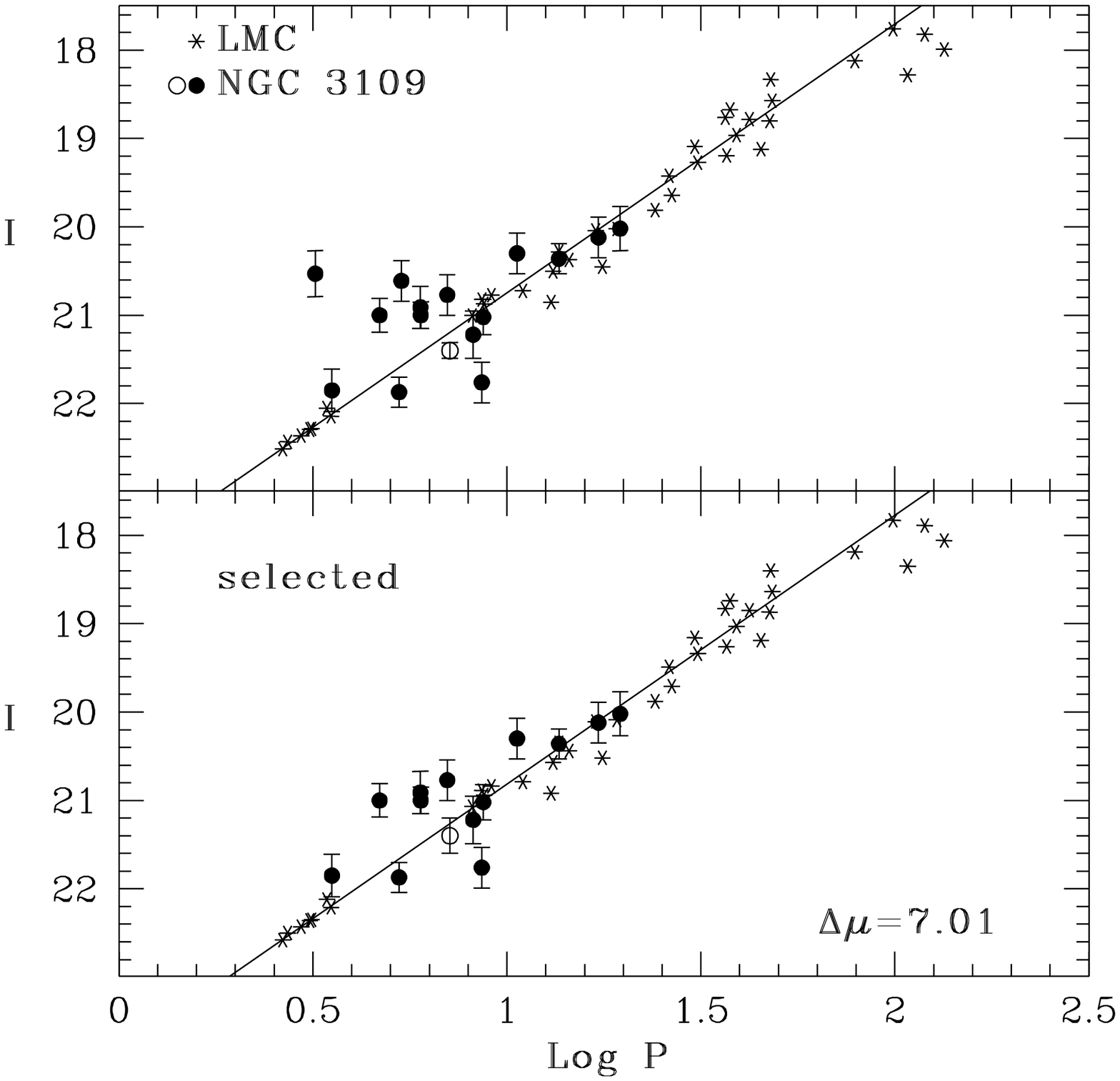} \caption{As in Fig. 4,
but for the $I$-band (the plotted errors are an upper limit, see Sect. 4
for details). The resulting apparent relative distance modulus is
$\Delta\mu_I=7.01 \pm 0.12$.} \label{fig7} \end{figure}

Figures \ref{fig4}, \ref{fig5}, \ref{fig6} and \ref{fig7} present the PL
relations from the mean $B$, $V$, $R$ and $I$ magnitudes of the Cepheids
listed in Table~\ref{table5}. {\it Filled circles} represent both SC and
our Cepheids in the field F1 of NGC~3109, while {\it open circles} refer
to the SC Cepheids located in the other fields. {\it Crosses} reproduce
the photoelectric data for the LMC Cepheids (Sandage 1988, Madore \&
Freedman, 1991), shifted by the zero point difference between the
NGC~3109 PL relation and the LMC one. The slopes of these two relations
appear similar in all four colors, in agreement with the postulate on
which the use of the Cepheids as distance indicators rests, i.e. that
their properties are independent of the host galaxy (\cf\ also
Section~7). For each color we built a PL relation with all the Cepheids
(upper plot) and one in which we used only the Cepheids with a well
defined light curve and a uniform phase coverage (referred to as
``selected'' from here on). This last selection allows us to include
only those Cepheids with the most reliable light curve parameters as
period, minimum, and mean magnitude.

In this paper, for consistency with CPB, we adopt a true distance
modulus to the LMC of 18.50 mag (van den Bergh, 1996) \footnote{We want
note that the current best estimate of the distance modulus of the LMC
is based on the light echo from the SN1987a ring. Panagia et al. (1997)
give $\mu_0=18.58\pm 0.03$ mag.} and a mean total extinction to the LMC
Cepheids $E(B-V)=0.08$~mag (\cf\  discussion in CPB). Note that Bessel
(1991) obtained a foreground reddening of the LMC which ranges from 0.04
to 0.09 mag, while the mean internal reddening is 0.06 mag, though in
some regions it reaches values as high as 0.3 mag. In this picture, it
is very difficult to estimate the mean total reddening of the sample of
the LMC Cepheids. As it will be discussed below, the adopted reddening
for the LMC has no effect on the determination of the relative (to the
LMC) and absolute distance modulus of NGC 3109: it only affects the
estimate of the total average reddening of its Cepheids.

In order to calculate the apparent distance moduli relative to the LMC in
all four bands, we obtained the best fit (least squares method) for the
PL relation of the LMC and then we calculated the best match of the PL
relation of NGC~3109 imposing the same slope of the PL relation of the
LMC. For an additional internal check of our results, we repeated these
calculations using different approaches.  First, we considered the two
different samples of Cepheids in NGC~3109 defined above. Moreover, while
the PL relation for the LMC was fitted with a straight line over the
range of periods $0< \log P<1.8$\footnote{Cepheids with $\log P>1.8$ are
excluded from the least-squares fit since both the evolutionary status
and the reddening of these longer period Cepheids are controversial.},
the two samples of NGC~3109 Cepheids were fitted both in the entire
range of periods and in the range $0.8<\log P<1.8$ in order to minimize
the effects of the Faint end selection bias in the small number data set
of the more distant galaxy NGC~3109. The resulting relative distance
moduli are all consistent within the errors. Thus we adopted the
relative distance moduli obtained with the Cepheids of the selected
sample of NGC~3109 without any constraint on $\log P$. In this way we
make use of the same range of periods for both the LMC and NGC~3109, but
we exclude those Cepheids with inaccurate parameters which might affect
the reliability of the fit (almost all of them are in the fainter part
of the PL relation). The resulting apparent distance moduli relative to
the LMC ($\Delta\mu$) are reported in the Table~\ref{table6}. The total
error on these apparent distance moduli is obtained adding in quadrature
the error of the fit and the mean internal photometric error (obtained
for each band as a mean of the errors in Table 2 up to magnitude 23.75).

\subsection{The absolute distance modulus} \label{absmod} In order to
determine the true distance modulus (TDM=$\mu_0$) to NGC~3109, the total
(foreground plus internal) extinction for the Cepheids must be evaluated.

\begin{deluxetable}{ccc} \tablenum{6} \tablecolumns{10} \tablewidth{0pc}
\tablecaption{Distance moduli relative to the LMC. \label{table6}}
\tablehead{ \colhead{Bandpass} & \colhead{$\Delta\mu$} & \colhead{Error}}
\startdata  B & 6.81 & 0.12\nl  V & 6.91 & 0.11\nl  R & 6.95 & 0.15\nl 
I & 7.01 & 0.15\nl \enddata \end{deluxetable}

Taking advantage of the multicolor apparent distance moduli (see
Table~\ref{table6}) and assuming that all of the wavelength dependence of
these moduli is due to the extinction, it is possible to fit all the data
simultaneously with an interstellar extinction law. In particular, we
assume that the reddening law for the LMC and NGC~3109 is the same as in
the Galaxy, and use the law derived by Cardelli et al. (1989, C89).

\begin{figure} \figurenum{8} \plotone{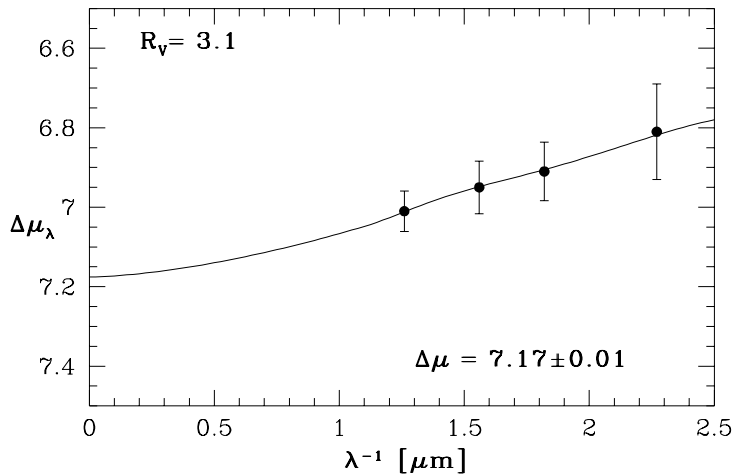} \caption{The NGC~3109
apparent distance moduli obtained in 4 bands are plotted as a function
of the corresponding inverse wavelength (in $\mu m^{-1}$). A weighted
least square fit of the Cardelli et al. (1989) extinction law ({\it
solid curve}) leads to a true relative distance modulus
$\Delta\mu_0=7.17\pm 0.01$.} \label{fig8} \end{figure}

Figure~\ref{fig8} shows the relative distance moduli plotted as a
function of the inverse wavelength characteristic of each band. As in
F88, the estimated errors were scaled as a function of the increasing
wavelength to reflect the decreasing strip width.  The solid line is the
weighted least square best fit of the adopted extinction law to the
moduli. Following C89, we have adopted $R_V = A_V/E(B-V)=3.1$, where
$A_V$ is the extinction in magnitudes in the $V$ band. With this
assumption, we obtained a reddening free relative distance modulus (RDM)
$\Delta\mu_0=7.17 \pm 0.01$ and a relative reddening $\Delta
E(B-V)=-0.09 \pm 0.02$ between the LMC and NGC 3109. This RDM
corresponds to a TDM $\mu_0=25.67$, adopting $\mu_0=18.50$ for the LMC.
A relative color excess $\Delta E(B-V)=-0.09$ would imply a negative
total mean reddening (TR) $E(B-V)=-0.01$ for NGC 3109, if we assume, as
discussed above and in CPB, $E(B-V)=0.08$ for the total reddening of the
LMC Cepheids. A negative reddening $E(B-V)=-0.07$ has been obtained by
Freedman \etal\ (1992) for NGC 300. Of course, a negative reddening has
no physical meaning, in terms of what we know about the interstellar
medium. First, we must note that in our case the uncertainty on the
$\Delta E(B-V)$ is of the order of 0.04 magnitudes (see below), and
therefore we can assume that the TR is compatible with zero. Still, we
would remain with the unlikely situation of a {\it zero} total
extinction for the Cepheids in NGC~3109. There might be many
explanations for such a result. Systematic calibration errors can play a
role in the determination of the relative extinction, though we can
exclude systematic errors larger than $\sim0.03$~mag. Another
interesting possibility is that the intrinsic color of the Cepheids in
NGC~3109 is systematically bluer than those in the fiducial sample of
the LMC, therefore leading to an underestimation of their reddening.
However, we believe that the largest uncertainty is still on the adopted
total reddening for the LMC. We have already discussed (previous
Section) the difficulties in estimating the mean internal and external
extinction toward the LMC Cepheids. According to Bessel's (1991)
results, the E(B-V)=0.08~mag for the LMC that we have adopted in this
series of papers might be too low. The comparison between the LMC and
NGC 3109 Cepheids seems to confirm that an $E(B-V)_{LMC}=0.1$~mag (see
also Freedman et al., 1991) or higher might be more appropriate.

The total error on the TDM of NGC~3109 is the combination of 1) the
error on the best fit of the extinction law of C89, 2) the uncertainty
on the absolute distance to the LMC, 3) the error on the photometric
zero point, and 4) the error due to the assumption of $R_V=3.1$.

The LMC best distance determination are based on the Cepheids ({\it cf.}
the compilation of Feast \& Walker, 1987), and the associated
uncertainty is $\pm 0.15$~mag; smaller errors seem too optimistic (see
van den Bergh 1996). This is still the dominant source of errors. The
uncertainty on the zero point calibration is $\pm 0.03$ mag (a mean of
the errors estimated in Sect. 2). Finally, C89 have shown that $R_V$ can
range from 2.75 to 5.3 within the Galaxy. Adopting this range in $R_V$
for NGC~3109, the reddening value and the relative distance modulus
would be in the interval $-0.01<E(B-V)<0.01$ and
$7.16<\Delta\mu_0<7.25$, respectively. In other words, a plausible error
estimates for the distance modulus as a consequence of the extinction
law uncertainties is $\pm 0.05$~mag.

In summary, we have that the true distance modulus of NGC~3109 is
$\mu_0=25.67 \pm 0.16$~mag and the total reddening is $E(B-V)=-0.01 \pm
0.04$~mag.

\subsection{Reddening--free PL Relation} In the previous Sections we
have used the information from several photometric bands and the fit of
an interstellar extinction law to the data in order to make an
extrapolation to infinitely long wavelengths and estimate the total
extinction and the true distance modulus.  However, even without solving
explicitly for the reddening, it is still possible to obtain an estimate
of the distance modulus which is independent of reddening. This can be
done by defining the reddening-free magnitude $W=V-R_V\times(B-V)$
(Freedman et al., 1992).

\begin{figure} \figurenum{9} \plotone{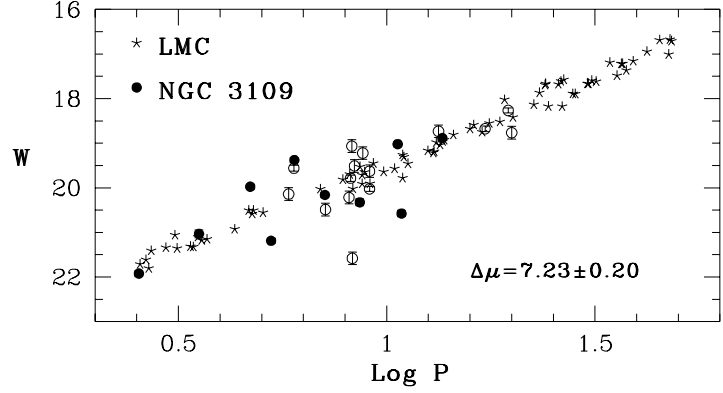} \caption{The
reddening-free $W-\log P$ relation for the NGC~3109 Cepheids ({\it open
circles} for the original SC variables and {\it filled circles} for the
new candidates), compared to the LMC Cepheids ({\it crosses}). The
corresponding true relative distance modulus is $\Delta \mu=7.23\pm
0.20$.} \label{fig9} \end{figure}

Figure~\ref{fig9} displays the $W-\log P$ relation for NGC~3109 (selected
sample) and for the LMC. From this relation we obtain a true relative
distance modulus $\Delta\mu_0=7.23\pm0.20$ which is consistent with the
true relative modulus obtained in the previous Section.

\begin{figure} \figurenum{10} \plotone{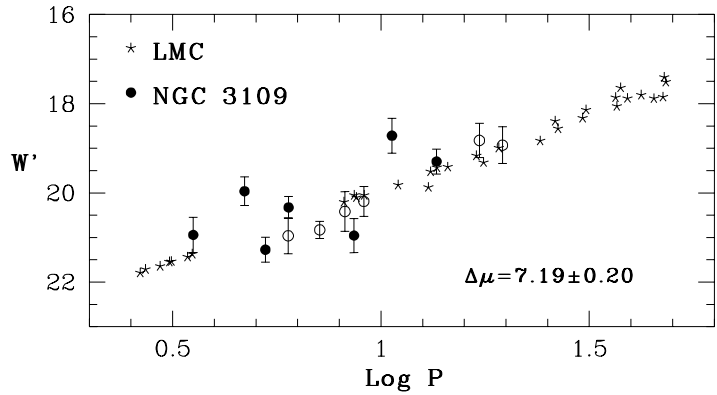} \caption {As in fig.
10, but for the new reddening-free $W'-\log P$ relation (see Sect. 5.3
for details) The corresponding true relative distance modulus is $\Delta
\mu=7.19\pm 0.20$.} \label{fig10} \end{figure}

The disadvantage of this reddening--free method is that the
uncertainties on the $W$ magnitudes are larger than the single errors on
the original $B$ and $V$ magnitudes, as $W$ results from a combination
of the two. For this reason we obtain an error on the true relative
modulus which is one order of magnitude larger than that on the modulus
obtained by the multiwavelength method. Moreover, $W$ depends on two of
the four available colors only, making the distance based on it
statistically less accurate. For this reason we have constructed a new
parameter $W'$ analog to the $W$, but depending on $R$ and $I$. The two
reddening free parameters ($W$ from $B$ and $V$ and $W'$ from $R$ and
$I$) are statistically uncorrelated and could be used as a further,
independent estimate of the distance modulus.  $W'$ was derived
generalizing the definition of $W$, and applying it to the $R$ and $I$
bands. In fact, we can substitute the parameter $R_V$ by ${\cal
R}(R,I)=A_I/E(R-I)$, and then define the new parameter $W'=I-{\cal
R}(I,R)*(R-I)$. ${\cal R}(I,R)$ can be easily derived by the C89 law.
Figure~\ref{fig10} displays the $W'-\log P$ relation for NGC~3109
(selected sample) and for the LMC. From this relation we obtain a true
relative distance modulus $\Delta\mu_0=7.19\pm0.20$ which is in a very
good agree with that obtained from the $W-\log P$ relation. This result
is a proof of the utility and reliability of this method. The use of
both $W$ and $W'$ can be used as a test of self consistency, and a way
of assessing external errors.

\section{The Period--Color relation} 
\begin{figure} \figurenum{11} \plotone{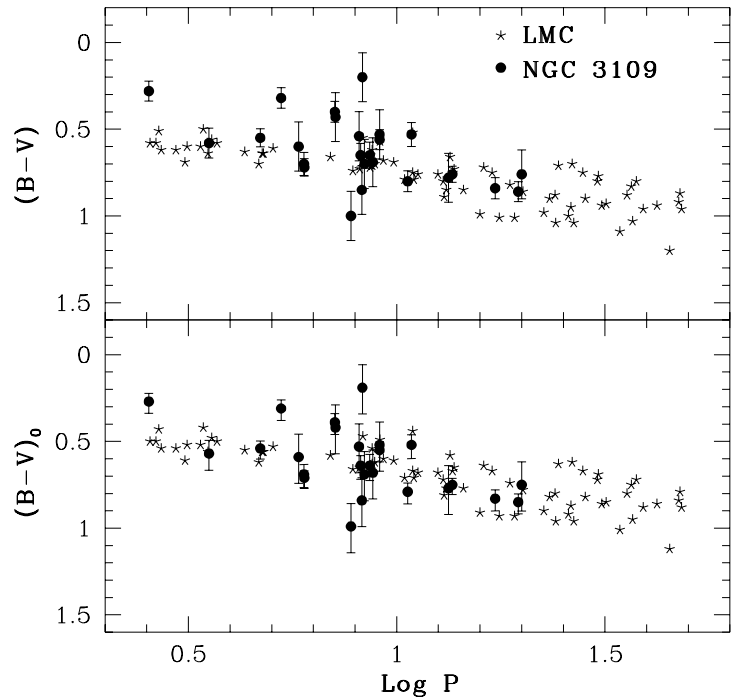} \caption{The $(B-V)$
colors for the Cepheids in the LMC ({\it crosses}) and in NGC~3109 ({\it
filled circles}) are plotted against the logarithm of the period in
days. For NGC~3109 we draw also the errors over the determination of the
colors. The {\it upper panel} shows the data without any correction for
the reddening, while the {\it lower panel} displays the absolute colors
[adopting E(B-V)=0.08 for the LMC]. For NGC~3109 we use E(B-V)=0.0 (see
text for details). The trends of the two populations is remarkably
similar. The larger scatter in NGC~3109 is due to the larger errors in
the determination of the mean magnitudes.} \label{fig11} \end{figure}

\begin{figure} \figurenum{12} \plotone{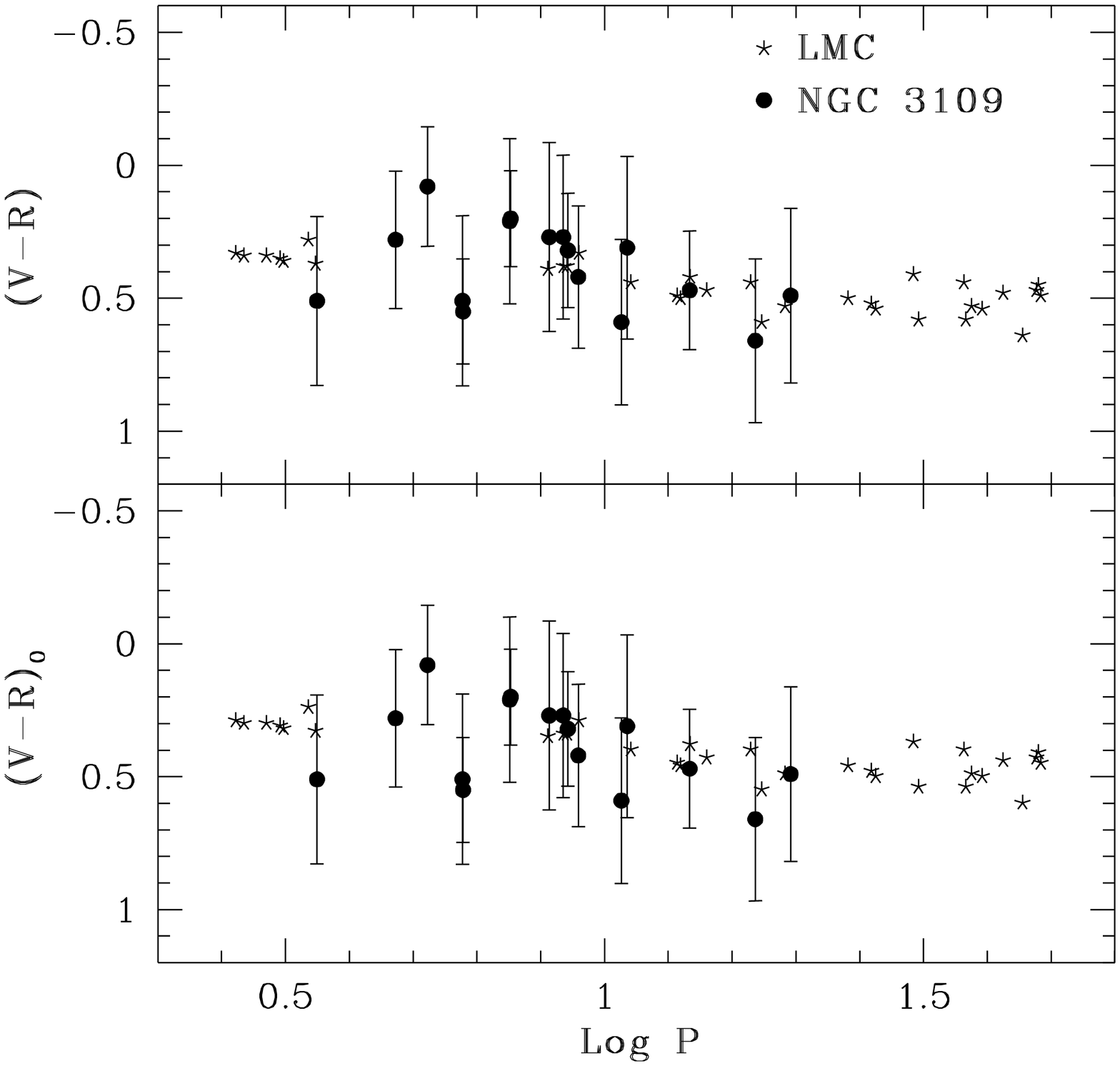} \caption{As in fig. 11,
but for the $(V-R)$ colors (the plotted errors are an upper limit, see
Sect. 4 for details).} \label{fig12} \end{figure}

\begin{figure} \figurenum{13} \plotone{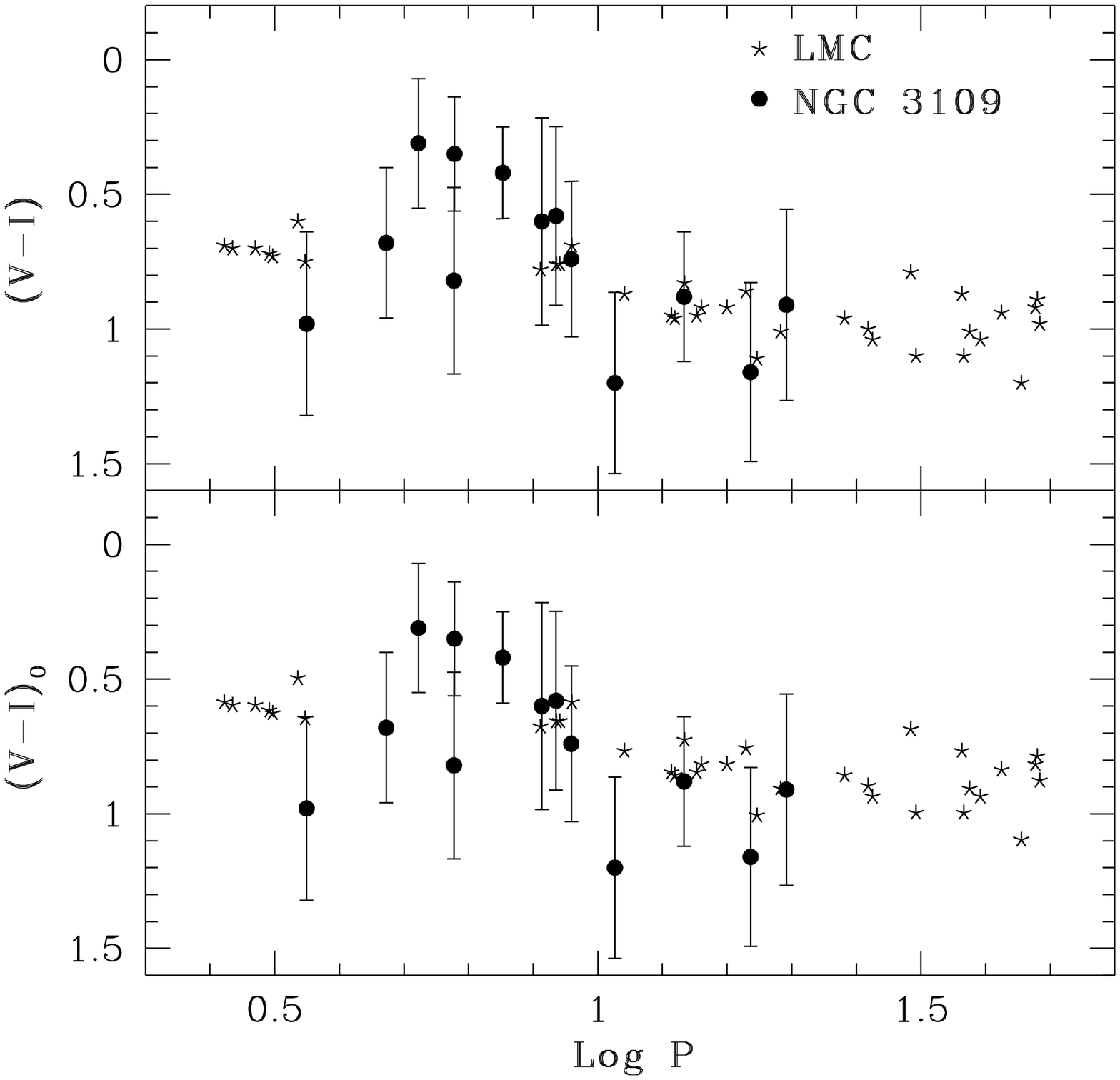} \caption{As in fig. 11,
but for the $(V-I)$ colors (the plotted errors are an upper limit, see
Sect. 4 for details).} \label{fig13} \end{figure}

\begin{figure} \figurenum{14} \plotone{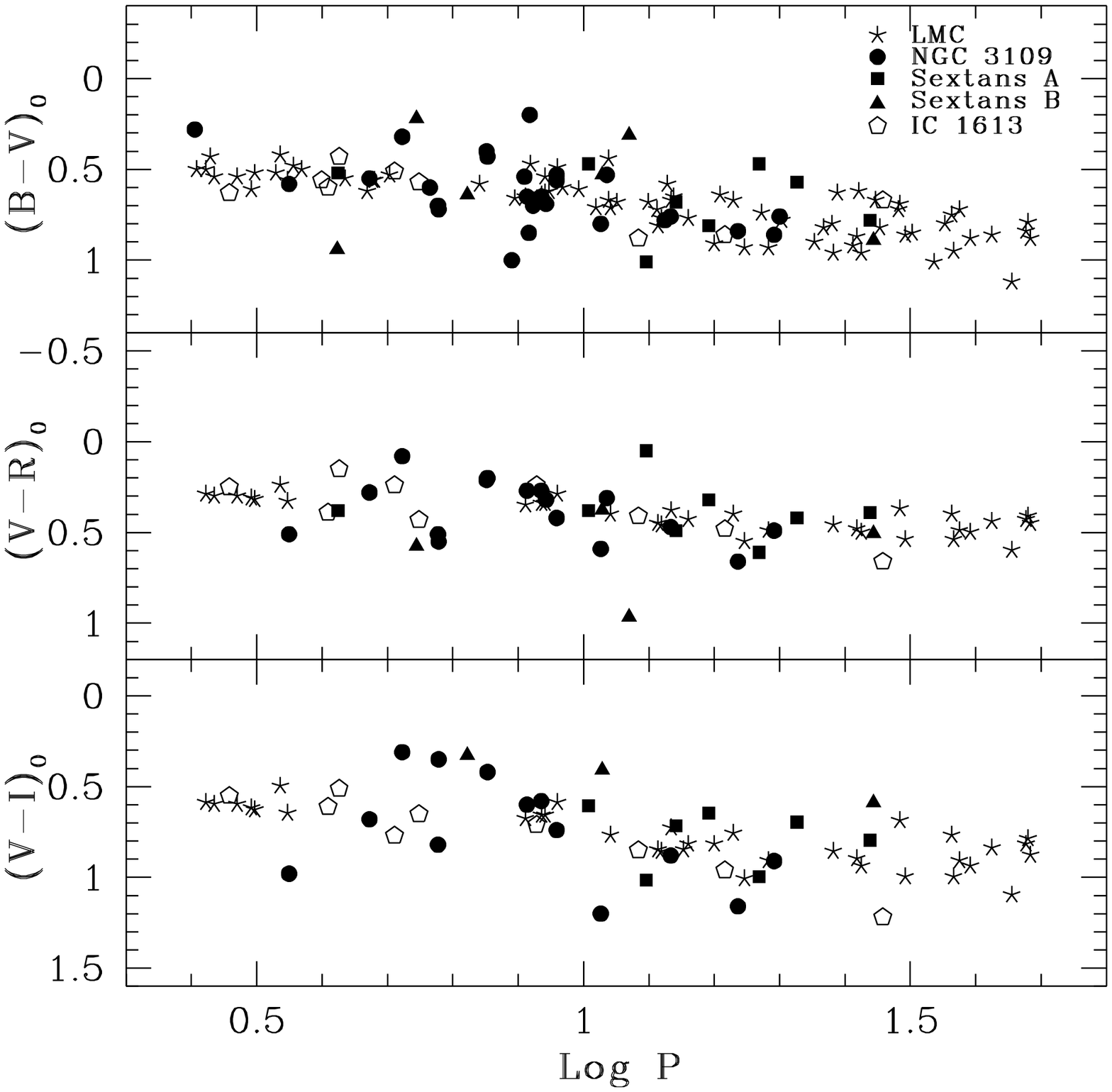} \caption{Period-Color
relation after the correction for reddening for the LMC, NGC~3109,
Sextant~A, Sextant~B and IC~1613. As for NGC~3109, the large scatter for
Sextant~A and Sextant~B is due to the larger errors in the determination
of the mean magnitudes.} \label{fig14} \end{figure}

In the Figs. \ref{fig11},
\ref{fig12}, and \ref{fig13}, we compare the Period-Color relations
($PC$) for the LMC and NGC~3109 for the three colors $(B-V)$, $(V-R)$
and $(V-I)$; for NGC~3109 the selected sample has been used (the errors
for the colors $(V-R)$ and $(V-I)$ are larger because we use the maximum
expected value and represent an upper limit, see Sect. 4 for details).
In each figure, the data in the {\it upper panel} are without reddening
correction and those in the {\it lower panel} are corrected according to
the reddening effects. The Fig. \ref{fig14} shows the same three
absorption-corrected PC relations for the LMC, NGC 3109, Sextans A
(PCP), Sextans B (PCP), and IC 1613 (F88). The slopes of these PC
relations are the same, within the uncertainties. The data of NGC~3109,
Sextans A, Sextans B, and IC 1613 have a larger scatter than for the
LMC, consistent with the lower accuracy in the determination of the mean
magnitudes.

\begin{figure} \figurenum{15} \plotone{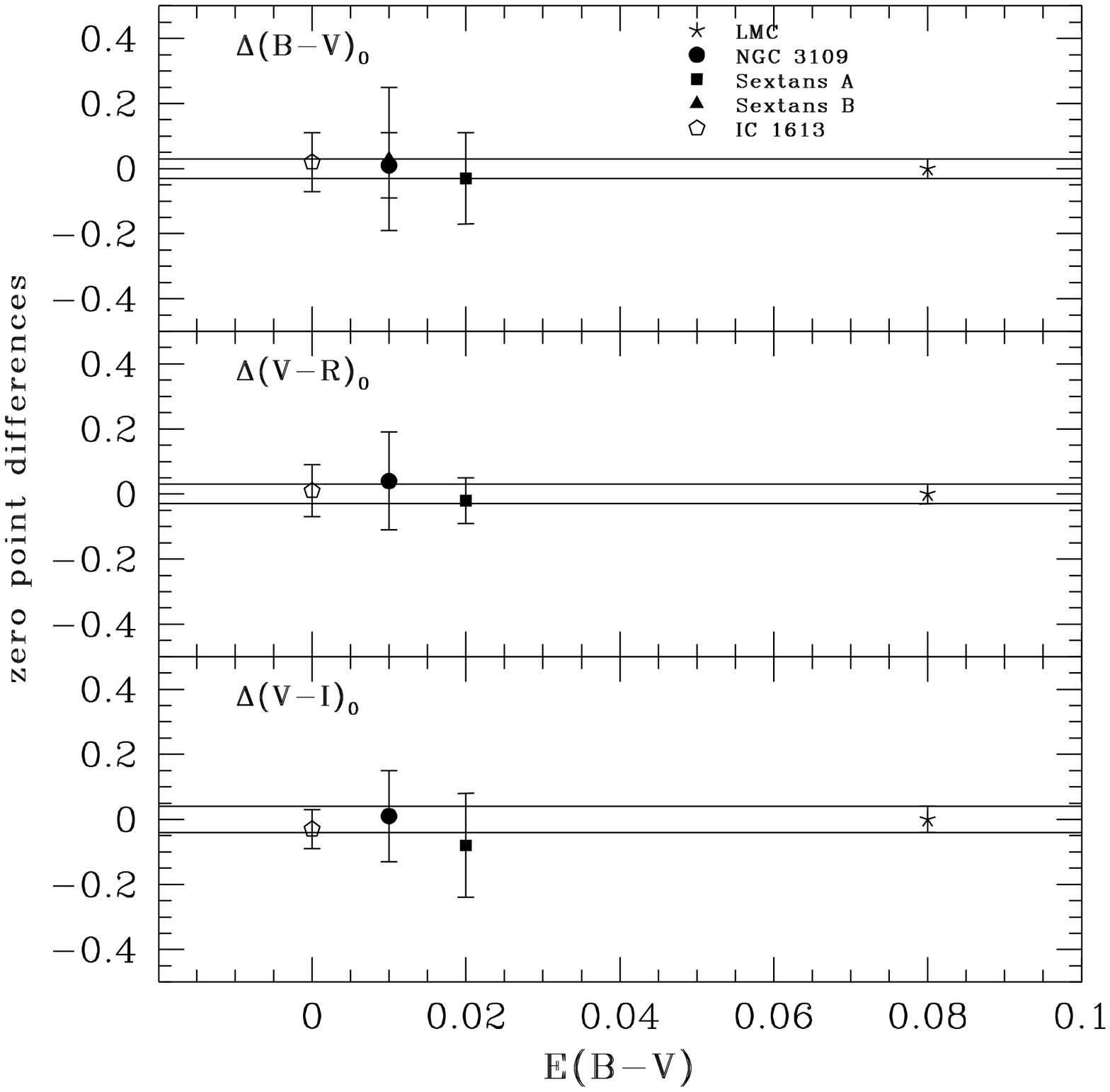} \caption{The zero point
difference (after reddening correction) between the $PC$ relations in
(B-V), (V-R), and (V-I) for the LMC and for NGC~3109, Sextans A, Sextans
B, and IC~1613 is plotted against the reddening. The {\it cross} is the
LMC and the two lines represent the range of errors for the LMC zero
point.} \label{fig15} \end{figure}

In order to obtain the zero point difference between the PC relations
for the LMC and the other galaxies, the LMC PC relations were fitted
with a straight line over the range of periods $0<\log P<1.8$, and a
linear fit of the PC relations was performed for the other galaxies
imposing the same slope of the LMC. In this way we could calculate the
mean color difference between the Cepheids in the LMC and in the other
four galaxies. Figure~\ref{fig15} shows that the mean color difference
after the reddening corrections is consistent with zero. This result
shows that, if we suppose that the colors of the Cepheids are independent
of the parent galaxy, the law by C89 allows to calculate relative
reddenings in a consistent way.

\begin{figure} \figurenum{16} \plotone{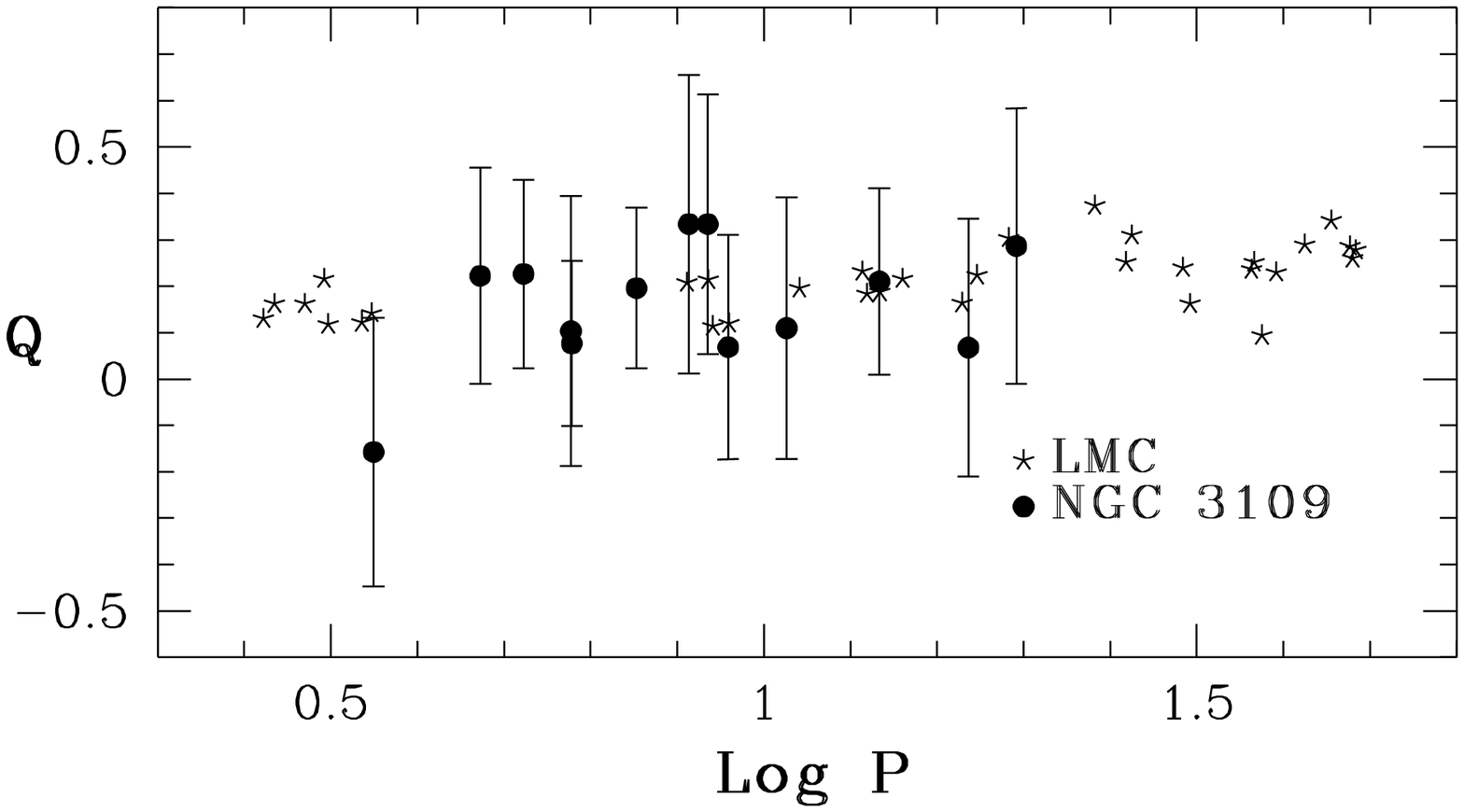} \caption{The
reddening--free parameter $Q=(B-V)-[E(V-I)/E(B-V)](V-I)$ is plotted as a
function of $Log P$ for both the selected sample of Cepheids in NGC~3109
({\it filled circles}) and in the LMC ({\it crosses}). The widths and
the slopes of the two distributions are the same as well as the zero
point. The larger scatter for the NGC~3109 Cepheids is consistent with
the larger photometric errors (for the magnitude $I$ the error
considered is an upper limit, see Sect. 4 for details).} \label{fig16}
\end{figure}

With the data set presented in this paper, we can check whether the
Cepheids in NGC~3109 have the same average colors as in the LMC. Figure
\ref{fig16} displays the reddening--free parameter
$Q=(B-V)-[E(V-I)/E(B-V)](V-I)$ as a function of $Log P$ for both the
selected sample of Cepheids in NGC~3109 ({\it filled circles}) and in the
LMC ({\it crosses}) (see Freedman \etal, 1992). The slopes of the two
distributions are the same as well as the zero point. The larger scatter
for the NGC~3109 Cepheids is consistent with the larger photometric
errors. The similarity of these two distributions suggest that the colors
of the Cepheids in these galaxies are the same.

\begin{figure} \figurenum{17} \plotone{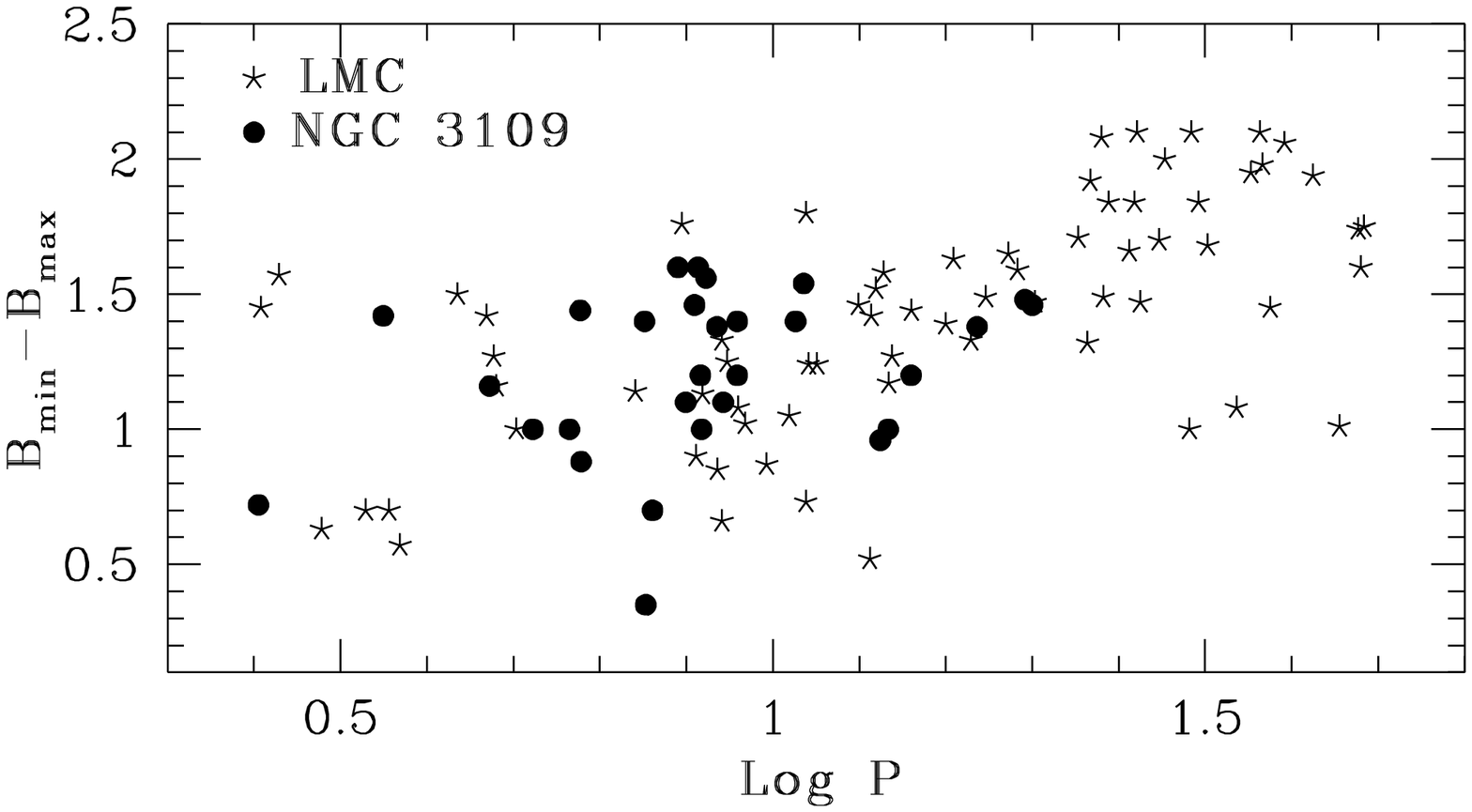} \caption{The light
curve amplitudes $(B_{min}-B_{max})$ for the LMC ({\it crosses}) and
NGC~3109 ({\it filled circles}) Cepheids are plotted against the
logarithm of the period in days. The trend of the two populations of
Cepheids is remarkably similar, suggesting similar properties.}
\label{fig17} \end{figure}

Finally, in Fig.~\ref{fig17} we compare the amplitude of the $B$ light
curves of the Cepheids in NGC~3109 and in the LMC. Also in this case we
obtain that the trend and the dispersion of the points for the two
populations of Cepheids are similar, suggesting similar properties.

\section{Conclusion}

In conclusion, an extended PL relation and the multicolor photometry
provide a new, more accurate, de-reddened distance of NGC 3109 relative
to the LMC. Adopting $\mu_0=18.50$  for the LMC, we obtain a true
distance modulus $\mu_0=25.67\pm 0.16$ for NGC 3109; if we assume a
reddening E(B-V)=0.08  for the LMC, the total reddening of the Cepheids
in NGC~3109 is $E(B-V)=-0.01\pm0.04$. The new reddening free distance
modulus corresponds to $1.36\pm0.10$ Mpc.

This new distance is consistent, within the errors, with the value
$\mu_0=25.5\pm 0.2$ given by CPB, though, formally, now NGC 3109 is
placed $\sim7\%$ further away. It is instructive to inspect the origin
of this difference. In CPB we had only B and V-band data. It was
difficult to apply the multiwavelength method to this data set, in view
of the closeness of the two bands and of the poor coverage in the V
band. For this reason, CPB preferred to adopt a mean reddening
E(B-V)=0.04 for NGC 3109 and estimate the true distance modulus with
this assumption in mind. The reddening estimate (from Burstein \&
Heiles, 1992) contour maps was admittedly poor (as noted also by BPC).
The difference between the reddening adopted by CPB and the present
direct estimate of the mean reddening toward NGC 3109 completely
explains the difference in the final absolute modulus. As further
evidence, we can refer to the exercise presented by Piotto et al.~(1995),
where we used the two B and V apparent relative distance moduli of NGC
3109 for a rough estimate of the true relative distance moduli with the
multiwavelength method applied also in the present paper. Even with a
poorer data set Piotto et al. (1995, cf. their Fig.~2) obtained a true
relative distance modulus $\Delta\mu_0=7.23\pm 0.10$, much closer to the
present estimate. This discussion shows by itself the importance of
extending the Cepheid light curves to the largest possible wavelength
interval. For this reason, we are observing these same Cepheids of
NGC~3109 in the JHK bands. We expect to reduce the error in the distance
and reddening estimate. For the importance of the JHK observations of
Cepheids, see Madore and Freedman (1991)

Finally, we want to briefly comment on the similarity of {\it (i)} the
slopes of the Cepheid PL and PC relations, {\it (ii)} the average color,
and {\it (iii)} of the properties of the amplitude-period relations which
result from a direct comparison of the data from five galaxies (cf.
discussion in Section 5.1, and 6). This is consistent with the
fundamental assumption on which the entire extragalactic distance scale
stands. However, we must also point out the large uncertainties still
present in the determination of all the above parameters (\cf\  also the
discussion in Section 5.2), in part as a consequence of the nature of
this kind of research, which requires a huge amount of telescope time.
It is clear that the dispersion (intrinsic and as a consequence of the
observational uncertainties) of the relations presented in Figs. 4-7,
and in Figs. 11-17 are too large for a firm conclusion on the
universality of the Cepheid properties. A lot of work on the theoretical
and observational side is still needed. In particular, we must take
advantage of the recent improvement in the near-IR detectors, where the
PL and PLC relations are much narrower.

\acknowledgements We wish to thank Wendy Freedman for providing us with
her LMC Cepheid data set in a computer readable form and Barry Madore
for the useful comments.\\ The authors acknowledge the support by the
Agenzia Spaziale Italiana. IM acknowledges the partial support of the
Istituto Italiano per gli Studi Filosofici, Napoli.

\end{document}